\documentclass[reprint,
 superscriptaddress,
 amsmath,amssymb,
 aps,
prl,
]{revtex4-2}

\usepackage[utf8]{inputenc} 
\usepackage[T1]{fontenc}    
\usepackage{hyperref}       
\usepackage{url}            
\usepackage{booktabs}       
\usepackage{amsfonts}       
\usepackage{nicefrac}       
\usepackage{microtype}      
\usepackage{graphicx}
\usepackage{xcolor, etoolbox}
\usepackage{amsmath,amsfonts,amsthm} 
\usepackage{mathtools}
\usepackage{ragged2e}
\usepackage{braket}      
\usepackage[english]{babel} 
\usepackage{enumitem}
\usepackage{xspace}
\usepackage{dsfont}
\usepackage{bm}
\usepackage{bbm}
\usepackage{dcolumn}
\usepackage{mathrsfs}

\newcommand{\Tr}{\textbf{Tr}}
\newcommand{\tr}{\text{Tr}}

\hypersetup{
    colorlinks,
    linkcolor=blue,
    citecolor=blue,
    urlcolor=blue,
    breaklinks=true 
}

\DeclarePairedDelimiter\abs{\lvert}{\rvert}

\begin{document}

\title{Area laws from classical entropies}

\author{Tobias Haas}
    \email{tobias.haas@ulb.be}
    \affiliation{Centre for Quantum Information and Communication, École polytechnique de Bruxelles, CP 165, Université libre de Bruxelles, 1050 Brussels, Belgium}

\begin{abstract}
The area law-like scaling of local quantum entropies is the central characteristic of the entanglement inherent in quantum fields, many-body systems, and spacetime. Whilst the area law is primarily associated with the entanglement structure of the underlying quantum state, we here show that it equally manifests in \textit{classical} entropies over measurement distributions when vacuum contributions dictated by the uncertainty principle are subtracted. Using the examples of the Gaussian ground and thermal states, but also the non-Gaussian particle state of a relativistic scalar field, we present analytical and numerical area laws for the entropies of various distributions and unveil how quantities of widespread interest such as the central charge and the (local) temperature are encoded in classical observables. With our approach, quantum entropies are no longer necessary to probe quantum phenomena, thereby rendering area laws and other quantum features directly accessible to theoretical models of high complexity as well as state-of-the-art experiments.
\end{abstract}

\maketitle

\textit{Introduction} --- The information content of a spatial subregion often scales with the area of its enclosing surface rather than its volume \cite{Bekenstein1973}. Subsumed under the term \textit{area law}, such scaling is typical for entanglement \cite{Horodecki2009} between spatially separated regions and occurs, in particular, for ground and low-lying states in field theories and many-body systems \cite{Calabrese2004,Calabrese2006,Hastings2007,Amico2008,Calabrese2009,Casini2009,Peschel2009,Eisert2010}, and black holes \cite{Hawking1975,Bombelli1986,Srednicki1993,Callan1994,Solodukhin2011}. The area law is commonly expressed via the entanglement entropy \cite{Plenio2007} and is therefore believed to rely on the knowledge of the local state. Since the local Hilbert space dimension scales exponentially with subsystem size, this poses major challenges for theoretical and experimental investigations when the number of degrees of freedom grows large. Consequently, the theoretical literature is dominated by results on Gaussian states and free theories that allow for an analytical treatment \cite{Calabrese2004,Calabrese2006,Hastings2007,Amico2008,Calabrese2009,Casini2009,Peschel2009,Eisert2010} (notable exceptions include, \textit{e.g.}, quasi-particles in scaling limits \cite{CastroAlvaredo2018a,CastroAlvaredo2018b,CastroAlvaredo2019a,CastroAlvaredo2019b}, perturbative interactions \cite{Hertzberg2013}, gauge theories \cite{Casini2014}, and matrix product states \cite{Schuch2008,Cirac2021}). For the same reason, experimental efforts have focused on finding efficient techniques to read out quantum entropies directly -- thereby bypassing quantum state tomography for sufficiently small system sizes \cite{Islam2015,Kaufman2016,Linke2018,Elben2018,Brydges2019} -- and on demonstrating the area law in Gaussian regimes \cite{Tajik2023}.

In this Letter, we argue that the area law is neither restricted to \textit{quantum} entropies, nor the \textit{full} density matrix. To this end, we consider functional phase-space descriptions of the underlying quantum state \cite{Wigner1931,Husimi1940,Schleich2001,Weedbrook2012,Mandel2013}, among which are the theoretically relevant Wigner $W$-distribution, as well as the experimentally accessible marginal and Husimi $Q$-distributions, and reveal area laws for their subtracted classical entropies and mutual informations. For Gaussian states, we show that the classical entropies of the former two are related to genuine quantum entropies and thus encode fundamental aspects of a quantum field theory such as its central charge. Remarkably, we find area laws also for non-Gaussian quasi-particle excitations and the Husimi $Q$-distribution. Our approach enables the assessment of the area law in terms of substantially less complex and experimentally accessible quantities. The feasibility of the herein-suggested methods is demonstrated in two companion papers \cite{PRL,PRA} concerned with area laws and local thermalization appearing in a spinor Bose-Einstein condensate after a quench. Technical details, also from the lattice perspective, are provided in a supplementary file \cite{SM}.

\textit{Notation} --- We use natural units $\hbar = k_{\text{B}} = 1$, denote quantum operators (classical variables) by bold (normal) letters $\boldsymbol{\rho}$ ($\gamma$), analogously for their traces $\Tr \{ \boldsymbol{\rho} \}$ ($\tr \{ \gamma \}$), and equip vacuum expressions with a bar $\bar{\gamma}$.

\textit{Quantum fields in phase space} --- We consider a relativistic scalar field theory in $1+1$ spacetime dimensions defined by the Hamiltonian
\begin{equation}
    \boldsymbol{H} = \frac{1}{2} \int \mathrm{d} x \, \left[ \boldsymbol{\pi}^2 + (\partial_x \boldsymbol{\phi})^2 + m^2 \boldsymbol{\phi}^2 \right],
    \label{eq:Hamiltonian}
\end{equation}
with a mass term $m$ and canonical commutation relations $[\boldsymbol{\phi} (x), \boldsymbol{\pi} (x')]= i \delta (x-x')$. We work in the Schrödinger picture \cite{Hatfield2018} where the field operator $\boldsymbol{\phi} (x)$ acts as a scalar and the conjugate momentum as a derivative $\boldsymbol{\pi} (x) = -i \delta_{\phi (x)}$ when applied to wave functionals $\Psi [\phi] = \braket{\phi | \psi}$, where $\phi (x)$ denotes the classical field configuration associated with the eigenstate $\ket{\phi}$, both defined via the eigenvalue equation $\boldsymbol{\phi} (x) \ket{\phi} = \phi (x) \ket{\phi}$.

The commutation relations dictate the geometry of phase space to be Euclidean \cite{Zhang1990}, thereby suggesting a Cartesian parameterization $\chi = (\phi, \pi)$ of phase-space distributions. Given some density matrix in the field basis $\rho[\phi, \phi'] = \braket{\phi | \boldsymbol{\rho} | \phi'}$, we define its functional Wigner $W$-distribution akin to quantum mechanics \cite{Zachos1999,Cembranos2021}
\begin{equation}
    \mathcal{W} [\chi] = \int \frac{\mathcal{D} \phi'}{\pi} \rho[\phi - \phi', \phi + \phi'] \, e^{2 i \int \mathrm{d} x \, \pi (x) \phi'(x)},
    \label{eq:WignerWDistribution}
\end{equation}
which is normalized to unity with respect to the functional integral measure $\int \mathcal{D} \chi \, \mathcal{W} [\chi] = \Tr \{ \boldsymbol{\rho} \} = 1$ (see \cite{SM} for a rigorous definition on the lattice). The Wigner $W$-distribution bears the same information content as the state itself and is of particular theoretical relevance for the simulation of the semi-classical dynamics of many-body systems via the well-known Truncated Wigner Approximation (TWA) \cite{Polkovnikov2010}. However, it is typically experimentally inaccessible in the many-body regime \cite{Lvovsky2009}.

Therefore, we will also be concerned with the \textit{directly} measurable marginal and Husimi $Q$-distributions \cite{PRL,PRA}. The former are defined as the diagonal elements of the state $\boldsymbol{\rho}$ in the corresponding eigenbases
\begin{equation}
    f [\phi] = \rho[\phi, \phi], \quad g [\pi] = \rho[\pi, \pi],
    \label{eq:MarginalDistributions}
\end{equation}
which equally follow from integrating out the complementary field in the Wigner $W$-distribution, \textit{i.e.}, $f [\phi] = \int \mathcal{D} \pi \, \mathcal{W} [\chi], g [\pi] = \int \mathcal{D} \phi \, \mathcal{W} [\chi]$, and with normalizations $\int \mathcal{D} \phi \, f [\phi] = \int \mathcal{D} \pi \, g [\pi] = 1$ understood. The latter is defined as the convolution of the Wigner $W$-distribution \eqref{eq:WignerWDistribution} with respect to the vacuum \footnote{In the quantum optics literature the Husimi $Q$-distribution is usually defined with an additional factor of $2 \pi$ \cite{Schleich2001}. However, all quantities of our interest are relative information measures and hence do not depend on the normalization}
\begin{equation}
    \mathcal{Q} [\chi] = \int \mathcal{D} \chi' \, \mathcal{W} [\chi'] \, \bar{\mathcal{W}} [\chi - \chi'],
    \label{eq:HusimiQDistribution}
\end{equation}
with normalization $\int \mathcal{D} \chi \, Q [\chi] = 1$. Here, $\bar{\mathcal{W}} [\chi]$ is Gaussian with block-diagonal covariance matrix in the sense that $\bar{\gamma}^{\mathcal{W}} = \bar{\gamma}^f \oplus \bar{\gamma}^g$ and $\bar{\gamma}^f (x,x') = \epsilon^2 \, \bar{\gamma}^g (x,x') = (\epsilon/2) \, \delta (x-x')$, where $\epsilon > 0$ denotes a lattice spacing such that $1/\epsilon$ acts as an ultraviolet regulator \footnote{At this point, the parameter $\epsilon$ also ensures correct mass dimensions}.

\begin{figure}[t!]
    \centering
    \includegraphics[width=0.99\columnwidth]{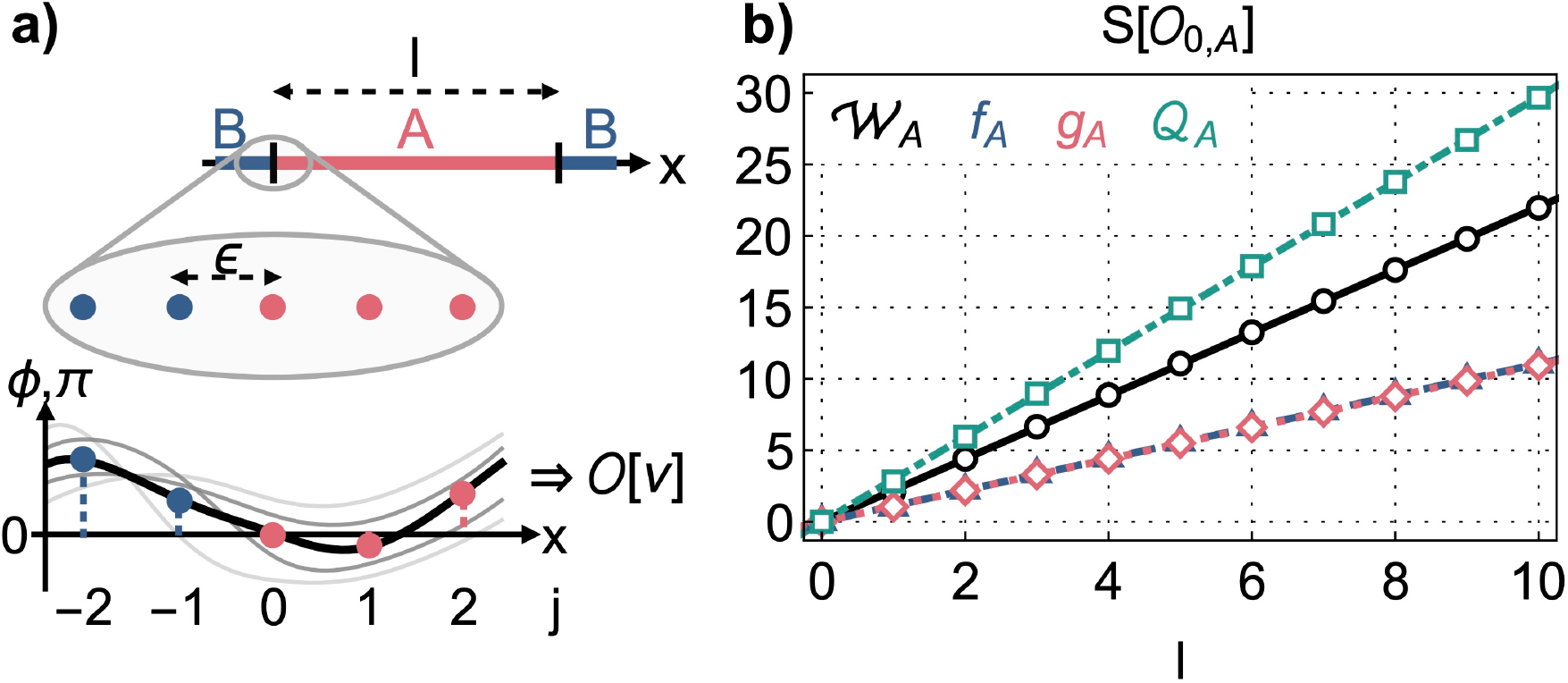}
    \caption{\textbf{a)} Illustration of subregion $A$ (red) and its complement $B$ (blue) in the continuum (lines) and on the lattice (points) where $x \to \epsilon j$ with lattice spacing $\epsilon >0$. At every (discrete) point in space, the distributions of our interest $\mathcal{O} [\nu]$ associate functional (quasi)-probability densities to all field configurations $\phi (x)$ and $\pi (x)$ (gray lines). \textbf{b)} Wigner (black), marginal (blue for $f$ and red for $g$), and Wehrl (petrol) entropies for the conformal ground state (points for $\epsilon=10^{-1},m=10^{-6}$, offsets subtracted in all figures) as functions of subsystem size $l$. All entropies are linear to leading order with proportionality constants close to their vacuum values $1 + \ln \pi, (1 + \ln \pi)/2, 1 + \ln 2 \pi$ (curves), respectively.}
    \label{fig:Setup}
\end{figure}

\textit{Typical local distributions} --- We are ultimately interested in the information structure of an interval $[0,l]$ which we refer to as subsystem $A$, see \autoref{fig:Setup} \textbf{a)}. Given any of the above classical distributions $\mathcal{O} = \mathcal{W}, f, g, \mathcal{Q}$ over their corresponding field configurations $\nu=\chi, \phi, \pi, \chi$, their local distributions are obtained by integrating out the complementary degrees of freedom, \textit{i.e.}, $\mathcal{O}_A [\nu_A] = \int \mathcal{D} \nu_B \, \mathcal{O} [\nu]$. Leaving dynamics aside, the set of typical states is built upon the ground state of the Hamiltonian \eqref{eq:Hamiltonian}, whose wave functional $\Psi_0 [\phi]$ is Gaussian. Additionally, we consider the thermal state $\boldsymbol{\rho}_T \sim \exp (- \boldsymbol{H}/T)$ of temperature $T > 0$ and a quasi-particle excitation of momentum $k$ with $\Psi_k [\phi] \sim \boldsymbol{a}^{\dagger}_k \Psi_0 [\phi]$ (all wave functionals and density matrices are given in \cite{SM}). Their corresponding local distributions are found after straightforward exercises in Gaussian integration
\begin{equation}
    \mathcal{O}_A [\nu_A] = \frac{1}{Z_A^{\mathcal{O}}} e^{- \frac{1}{2} \int_A \mathrm{d}x  \mathrm{d}x' \nu_A^T (x) (\gamma_A^{\mathcal{O}})^{-1} (x,x') \nu_A (x')} \times \kappa_A^{\mathcal{O}} [\nu_A],
    \label{eq:LocalDistributionGeneral}
\end{equation}
with the quadratic form
\begin{equation}
    \kappa_A^{\mathcal{O}} [\nu_A] = \lambda_A^{\mathcal{O}} + \int_A \mathrm{d}x \, \mathrm{d}x' \nu_A^T (x) (\Lambda_A^{\mathcal{O}})^{-1} (x,x') \nu_A (x').
\end{equation}
Therein, $\gamma_A^{\mathcal{O}} (x,x') = \int_A \mathcal{D}\nu_A \mathcal{D} \nu'_A \mathcal{O}_A \nu_A (x) \nu'_A(x')$ denotes the local covariance matrix which encodes the well-known equal-time two-point correlation functions of the scalar field and $Z^{\mathcal{O}}_A = \text{det}^{1/2} (2 \pi \gamma^{\mathcal{O}}_A)$ is the normalization of the Gaussian part. The quantities $\lambda_A^{\mathcal{O}}$ and $\Lambda_A^{\mathcal{O}}$ describe non-Gaussian terms and are involved functionals of the global covariance matrix $\gamma^{\mathcal{O}}$, see \cite{SM} for details.

\textit{Subtracted classical entropies and classical mutual informations} --- With the local distributions \eqref{eq:LocalDistributionGeneral} we associate the class of classical Rényi entropies in a functional sense
\begin{equation}
    S_{r} [\mathcal{O}_A] = \frac{1}{1 - r} \ln \left( \int \mathcal{D} \nu_A \, \mathcal{O}_A^{r} \right),
    \label{eq:ClassicalRenyiEntropy}
\end{equation}
where $r \in (0,1) \cup (1,\infty)$ specifies the entropic order, such that the classical entropy $S [\mathcal{O}_A] = - \int \mathcal{D} \nu_A \, \mathcal{O}_A \ln \mathcal{O}_A$ is recovered in the limit $r \to 1$ \footnote{Both definitions assume $\mathcal{O}_A [\nu_A] \ge 0 \, \forall \nu_A$, which is always fulfilled for the marginal and Husimi $Q$-distributions. Although the global Wigner $W$-distribution is negative already for all pure non-Gaussian states by Hudson's theorem \cite{Hudson1974}, its local distribution is very likely to be positive when the subsystem's size is small compared to the system's total size, as then the underlying state is highly mixed, which follows from Page's theorem \cite{Page1993}. We have checked the positivity of all distributions underlying Figure 2 numerically.}. In contrast to their quantum analogs, such entropies do generally \textit{not} measure the \textit{mixedness} of the underlying quantum state, but rather the \textit{localization} of the considered distribution over the classical field configurations \footnote{We refer to \cite{VanHerstraeten2021a,VanHerstraeten2021b,Haas2022c,Haas2022d,Haas2024} for a treatise on differential entropies from the perspective of continuous majorization theory}. Since the fields $\phi (x)$ and $\pi (x)$ are incompatible, their entropies are subjected to the uncertainty principle in the form of entropic uncertainty relations (see \cite{Coles2017,Hertz2019} for reviews). The common feature of the plethora of such relations (see, \textit{e.g.}, \cite{Bialynicki-Birula1975,Wehrl1978,Wehrl1979,Lieb1978,Lieb2014,Haas2021b,VanHerstraeten2021a,VanHerstraeten2021b,Haas2024}) is that any phase-space entropy is minimized by its vacuum expression \footnote{For the marginal entropies this only holds when considering their sum, while for the Wigner this is an open conjcture, see \cite{VanHerstraeten2021a}}. Together with the additivity of the entropy for the product-form vacuum, this implies 
\begin{equation}
    S_{r} [\mathcal{O}_A] \ge S_{r} [\bar{\mathcal{O}}_A] \sim l/\epsilon.
    \label{eq:EUR}
\end{equation}
The lower bound is extensive, thereby demonstrating a state-independent volume-law-like scaling of any classical Rényi entropy to leading order, see \autoref{fig:Setup} \textbf{b)} for the ground state. Further, classical Rényi entropies diverge even globally in the continuum limit, showing that \textit{relative} entropic measures should be preferred for describing entropic uncertainty of quantum fields \cite{Haas2021a,Haas2022b,Haas2023}.

Motivated by these observations, we wish to describe the entanglement of subsystem $A$ by assessing the next-to-leading order behavior of \eqref{eq:ClassicalRenyiEntropy}. To this end, we introduce the \textit{subtracted} classical Rényi entropy 
\begin{equation}
    \Delta S_{r} [\mathcal{O}_A] \equiv S_{r} [\mathcal{O}_A] - S_{r} [\bar{\mathcal{O}}_A],
    \label{eq:SubtractedClassicalRenyiEntropy}
\end{equation}
which measures the uncertainty deficit of some generic distribution $\mathcal{O}_A$ with respect to the vacuum $\bar{\mathcal{O}}_A$. Besides characterizing entanglement, we are also interested in quantifying correlations -- quantum \textit{and} classical -- among subsystem $A$ and its complement $B$, for which we define the classical Rényi mutual information
\begin{equation}
    I_r [\mathcal{O}_A : \mathcal{O}_B] = S_r [\mathcal{O}_A] + S_r [\mathcal{O}_B] - S_r [\mathcal{O}].
    \label{eq:ClassicalRenyiMutualInformation}
\end{equation}
For $r \to 1$, this serves as a measure for the total correlations being zero if and only if $A$ and $B$ are uncorrelated, that is, if the global distribution is of product form $\mathcal{O} [\nu] = \mathcal{O}_A [\nu_A] \times \mathcal{O}_B [\nu_B]$ \footnote{As we will see below, non-negativity of \eqref{eq:ClassicalRenyiMutualInformation} is ensured for all entropic orders $r$ in the case of Gaussian distributions}. Being defined as a relative measure, extensive contributions cancel out naturally.

\begin{figure*}[t!]
    \centering
    \includegraphics[width=0.98\textwidth]{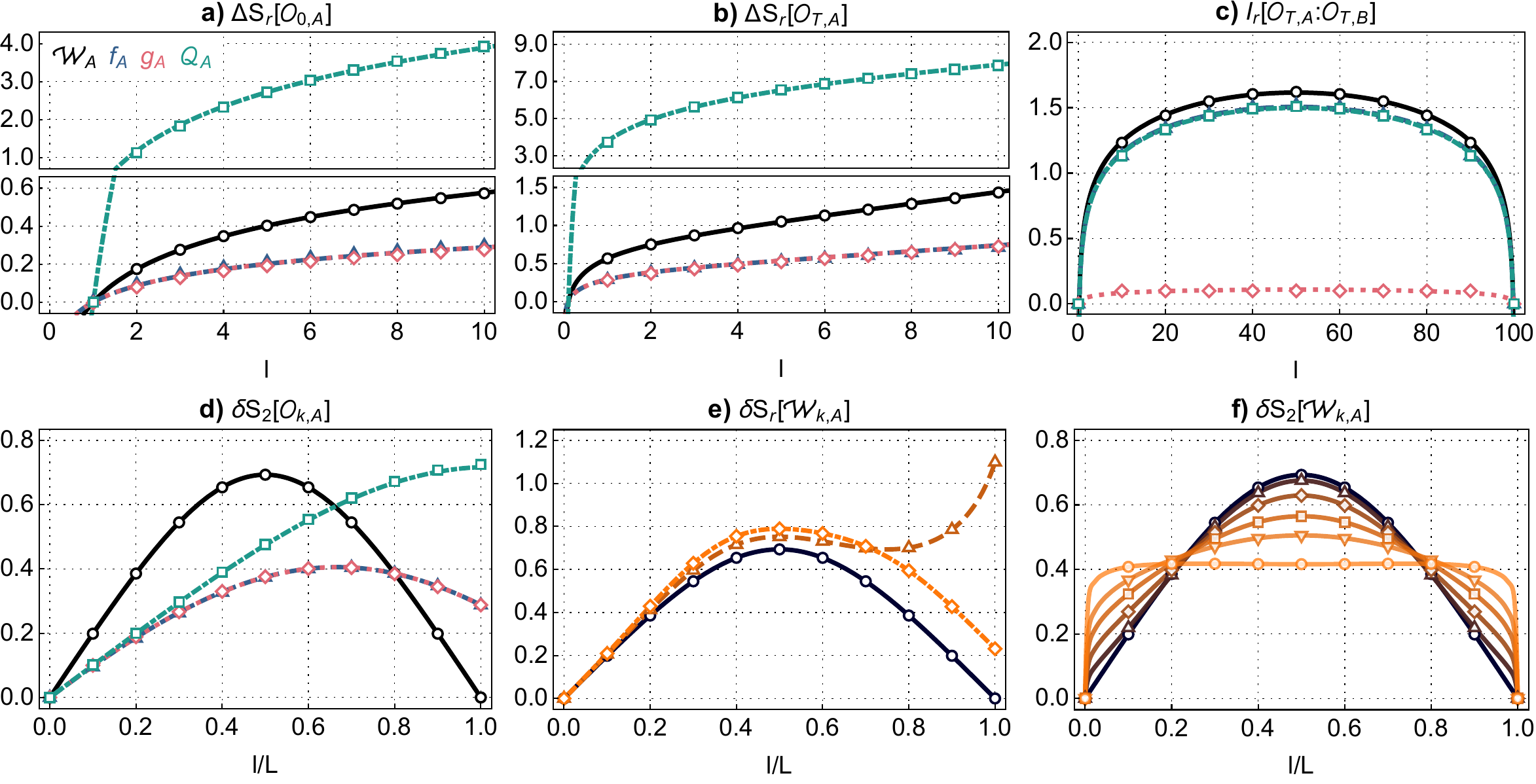}
    \caption{\textbf{a)} Subtracted classical Rényi entropies of the conformal ground state. The analytical formulae \eqref{eq:SubtractedClassicalWignerAndMarginalEntropiesGroundState} for the Wigner $W$- (black) and the marginal (blue and red) distributions are in excellent agreement with lattice results where $\epsilon = 10^{-1}, m = 10^{-6}, L=10^2 \gg l=10$ (points). Numerically, we also find an area law for the subtracted Wehrl entropy $\Delta S_r [\mathcal{Q}_{A}] - (l/\epsilon) \ln 2$ (petrol points), with prefactor $\approx 7/4$ extracted from a fit (petrol curve), see \cite{SM}. \textbf{b)} Analogous analysis for the thermal state of temperature $T = 1/l$ and an effective temperature $ \approx 2.87 T$ for the subtracted Wehrl entropy. \textbf{c)} Classical mutual informations for the same scenario. Here, all curves correspond to finite-size area law fits $(a/4) \ln [L/(\pi l) \sin (\pi l /L)]$. \textbf{d)} Uncertainty surplus $\delta S_2 [\mathcal{O}_{k,A}]$ of order two for a quasi-particle with $p=\pi/\epsilon, m=10$ in a box of length $L=10$. The analytical formula \eqref{eq:SubtractedRenyiEntropiesParticle} (curves) is supported by lattice calculations (points). \textbf{e)} Similar analysis for the Wigner $W$-distribution and varying entropic orders $r=2$ (black), $r=3$ (brown), $r=4$ (orange), see \cite{SM} for explicit expressions. \textbf{f)} Transition of the Rényi-2 entanglement entropy $\delta S_2 [\mathcal{W}_{k,A}]$ for decreasing masses $m=10,1,0.5,0.3,0.2,0.1$ (black to orange, curves are interpolations).}
    \label{fig:Entropies}
\end{figure*}

Computing both quantities of our interest for distributions of the type \eqref{eq:LocalDistributionGeneral} boils down to yet another exercise in Gaussian integration. The integral $\int \mathcal{D} \nu_A \, \mathcal{O}_A^r$ is solved by exploiting the central identity of quantum field theory, that is, pulling $\kappa_A^{\mathcal{O}}$ out of the integral, \textit{i.e.}, \footnote{Note here that depending on whether $\mathcal{O}_A$ covers phase space or is a marginal distribution, we have $\dim (\gamma_A^{\mathcal{O}}) = 2 \, l / \epsilon$ or $\dim (\gamma_A^{\mathcal{O}}) = l / \epsilon$, respectively}
\begin{equation}
    \int \mathcal{D} \nu_A \, \mathcal{O}_A^r = \sqrt{\frac{\text{det}^{1-r} (2 \pi \gamma_A^{\mathcal{O}})}{r^{(2) l / \epsilon}}} \, U_A^{\mathcal{O}} (r),
    \label{eq:LocalRenyiPower}
\end{equation}
with the non-Gaussian contribution
\begin{equation}
    U_A^{\mathcal{O}} (r) = (\kappa_A^{\mathcal{O}} [\partial_{\zeta}])^r e^{\frac{1}{2r} \int \mathrm{d}x \mathrm{d}x' \, \zeta^T (x) \gamma_A^{\mathcal{O}} (x,x') \zeta (x')} \big \vert_{\zeta = 0}.
\end{equation}
For integer $r$, using the binomial and Isserlis-Wick theorems reduces the latter to a set of contractions of $\gamma_A^{\mathcal{O}} \Lambda_A^{\mathcal{O}}$ and its transpose, which can then be evaluated with a diagrammatic technique reminiscent of Feynman diagrams, see \cite{SM}. Thus, the classical Rényi entropy reads
\begin{equation}
    S_r [\mathcal{O}_A] = \frac{1}{2} \ln \det (2 \pi \gamma_A^{\mathcal{O}}) + \frac{1(2)}{2} \frac{\ln r}{r-1} \frac{l}{\epsilon} + \delta S_r [\mathcal{O}_A],
\end{equation}
where the second term unveils the extensive scaling of every classical entropy and the third term $\delta S_r [\mathcal{O}_A] = \ln (U_A^{\mathcal{O}})/(1-r)$ accounts for non-Gaussian contributions. Since the vacuum is of Gaussian form, its entropy is fully specified by the vacuum covariance $\bar{\gamma}^{\mathcal{O}}_A$. Importantly, extensive scalings stemming from the uncertainty principle cancel out for both quantities of our interest, such that
\begin{equation}
    \begin{split}
        \Delta S_r [\mathcal{O}_A] &= \frac{1}{2} \ln \det \left[ \gamma_A^{\mathcal{O}} (\bar{\gamma}_A^{\mathcal{O}})^{-1} \right] + \delta S_r [\mathcal{O}_A], \\
        I_r [\mathcal{O}_A : \mathcal{O}_B] &= \frac{1}{2} \ln \frac{\det (\gamma_A^{\mathcal{O}} \gamma_B^{\mathcal{O}})}{\det (\gamma^{\mathcal{O}})} + \delta I_r [\mathcal{O}_A : \mathcal{O}_B].
    \end{split}
    \label{eq:SubtractedClassicalRenyiEntropyAndClassicalRenyiMutualInformationFormula}
\end{equation}
with $ \delta I_r [\mathcal{O}_A : \mathcal{O}_B] = \delta S_r [\mathcal{O}_A] + \delta S_r [\mathcal{O}_B] - \delta S_r [\mathcal{O}]$.

\textit{Gaussian states} --- Remarkably, both the subtracted classical Rényi entropy and mutual information become \textit{independent} of the entropic order $r$ when considering Gaussian states. This underpins a striking relation to a quantum entropy which can be found for the Wigner $W$-distribution and its marginals. To this end, we recall the purity of a Gaussian state $\mu_A = \Tr \{ \boldsymbol{\rho}_A^2 \} = \text{det}^{-1/2} \left(2 \gamma_A^{\mathcal{W}} \right)$ \cite{Serafini2017}, and note that the Wigner covariance matrix contains the two-point correlation functions of the local state $\boldsymbol{\rho}_A$ by definition. This leads to the relations
\begin{equation}
    \hspace{-0.15cm}\Delta S_r [\mathcal{W}_A] = S_2 [\boldsymbol{\rho}_A], \,\,\,\, I_r [\mathcal{W}_A : \mathcal{W}_B] = I_2 [\boldsymbol{\rho}_A : \boldsymbol{\rho}_B],
    \label{eq:SubtractedClassicalRenyiWignerEntropyAndMutualInformationGaussianRelation}
\end{equation}
where $S_2 [\boldsymbol{\rho}_A] = - \ln \mu_A$ and $I_2 [\boldsymbol{\rho}_A : \boldsymbol{\rho}_B] = S_2 [\boldsymbol{\rho}_A] + S_2 [\boldsymbol{\rho}_B] - S_2 [\boldsymbol{\rho}]$ denote the Rényi-2 entanglement entropy and the Rényi-2 mutual information, respectively, revealing that quantum entanglement measures reduce to classical uncertainty measures for Gaussian Wigner $W$-distributions \footnote{The relations are generalizations of two relations found for finitely many bosonic modes when $r \to 1$, see \cite{Adesso2012}}. When the latter is of product form $\mathcal{W}_A [\chi_A] = f_A [\phi_A] \times g_A [\pi_A]$ -- which includes both the ground and the thermal state -- the two relations in \eqref{eq:SubtractedClassicalRenyiWignerEntropyAndMutualInformationGaussianRelation} extend to the marginal distributions by additivity, to wit 
\begin{equation}
    \begin{split}
        \Delta S_r [f_A] + \Delta S_r [g_A] &= S_2 [\boldsymbol{\rho}_A], \\
        I_r [f_A : f_B] + I_r [g_A : g_B] &= I_2 [\boldsymbol{\rho}_A : \boldsymbol{\rho}_B].
    \end{split}
    \label{eq:SubtractedClassicalMarginalWignerEntropyAndMutualInformationGaussianRelation}
\end{equation}

Let us now assess the ground-state entanglement of the Hamiltonian \eqref{eq:Hamiltonian}. Quantum correlations between $A$ and $B$ are exponentially suppressed beyond the correlation length $\xi$, which is given by the inverse mass $\xi = 1/m$ for the massive theory $l \gg 1/m$ and by the subregion size $\xi = l$ in the conformal limit $l \ll 1/m$. Upon using the well-known expression for the Rényi-2 entanglement entropy \cite{Calabrese2004}, we find the area laws
\begin{equation}
    \Delta S_r [\mathcal{W}_{0,A}] = \Delta S_r [f_{0,A}] + \Delta S_r [g_{0,A}] = \frac{c}{4} \ln \left(\frac{\xi}{\epsilon} \right),
    \label{eq:SubtractedClassicalWignerAndMarginalEntropiesGroundState}
\end{equation}
with the central charge $c=1$ for the scalar field. These analytic formulae (curves) are supported by numerical results on the lattice (points), which are compared in \autoref{fig:Entropies} \textbf{a)} for the conformal case. The lattice computations indicate that the entanglement is evenly distributed over field and momentum field entropies in the continuum, \textit{i.e.}, $\Delta S_r [f_{0,A}] = \Delta S_r [g_{0,A}] = (c/8) \ln (\xi / \epsilon)$ \footnote{For finite $\epsilon > 0$, both marginal entropies keep the area-law form \eqref{eq:SubtractedClassicalWignerAndMarginalEntropiesGroundState} with the coefficient of $\Delta S_r [f_{0,A}]$ being slightly larger than the one of $\Delta S_r [g_{0,A}]$, see \cite{SM}}, revealing another fundamental insight: the area law does \textit{not} rely on the full information encoded in the local density matrix -- it is present already for the density matrix' diagonal elements in the field bases. Further, we find the so-called subtracted Wehrl entropy associated with the Husimi $Q$-distribution to fulfill an area law as well when we additionally subtract $(l/\epsilon) \ln 2$, which corresponds to the entropy gain of the convolution ($\gamma^{\mathcal{Q}}_A = \gamma^{\mathcal{W}}_A + \bar{\gamma}_A^{\mathcal{W}}$). The prefactor $\approx 7/4$ is numerically determined, see \cite{SM}.

We proceed with the thermal state. With the corresponding result for the Rényi-2 entanglement entropy \cite{Calabrese2004} we obtain the finite-temperature area laws
\begin{equation}
    \begin{split}
        \Delta S_r [\mathcal{W}_{T,A}] &= \Delta S_r [f_{T,A}] + \Delta S_r [g_{T,A}] \\
        &= \frac{c}{4} \ln \left[ \frac{\sinh ( \pi T \xi )}{\pi \epsilon T} \right],
    \end{split}    \label{eq:SubtractedClassicalWignerAndMarginalEntropiesThermalState}
\end{equation}
see \autoref{fig:Entropies} \textbf{b)}. The subtracted Wehrl entropy obeys the same scaling with an effectively higher temperature $\approx 2.87T$ since $\bar{\gamma}_A^{\mathcal{W}}$ is diagonal also in momentum space, thereby increasing only the zero-mode populations. While subtracted classical entropies become linear $ \Delta S_r [\mathcal{O}_{T,A}] \propto \xi/T$ for large temperatures $T \gg 1/\xi$, the area law persists for classical mutual informations, see \autoref{fig:Entropies} \textbf{c)}. More generally, the area law holds for the thermal state of every \textit{local} (not necessarily free) theory, \textit{i.e.}, $I_r [\mathcal{O}_{T,A} : \mathcal{O}_{T,B}] \le a \abs{\partial A}$ for some $a>0$, see \cite{SM} for a proof. When considered for a finite volume $L$, we find most of the correlations to be contained in the field $\phi$, since $\Delta S_r [f_{T,A}]$ falls off faster than $\Delta S_r [g_{T,A}]$ for $l \to L$, \textit{i.e.}, when $A$ covers most of the total system, see \cite{SM} for details. As the convolution mainly erases $\pi$-correlations, this implies $I_r [\mathcal{Q}_{T,A} : \mathcal{Q}_{T,B}] \to I_r [f_{T,A} : f_{T,B}]$ when $\epsilon \to 0$. 

\textit{Non-Gaussian states} --- Beyond Gaussian states, classical and quantum entropies of order two are related via the Wigner-Weyl transformation. The expectation value of some operator $\boldsymbol{T}_A$ can be calculated in phase space using the trace formula $\Tr \{ \boldsymbol{\rho}_A \boldsymbol{T}_A \} = \int \mathcal{D} \nu_A 2 \pi \, \mathcal{W}_A [\nu_A] \, \mathcal{T}_A [\nu_A]$, where $\mathcal{T}_A [\nu_A]$ is the Wigner symbol of $\boldsymbol{T}_A$ \cite{Serafini2017}. Since the Wigner symbol of $\boldsymbol{\rho}_A$ is $\mathcal{W}_A$, \textit{every} state obeys
\begin{equation}
    \hspace{-0.15cm}\Delta S_2 [\mathcal{W}_A] = S_2 [\boldsymbol{\rho}_A], \,\,\,\, I_2 [\mathcal{W}_A : \mathcal{W}_B] = I_2 [\boldsymbol{\rho}_A : \boldsymbol{\rho}_B],
    \label{eq:SubtractedClassicalRenyi2WignerEntropyAndMutualInformationGeneralRelation}
\end{equation}
showing that subtracted Wigner and entanglement entropies and mutual informations of order two \textit{agree}. Besides, every classical mutual information is a lower bound to the quantum mutual information $I [\mathcal{O}_A : \mathcal{O}_B] \le I [\boldsymbol{\rho}_A : \boldsymbol{\rho}_B]$, which is a consequence of the stronger subadditivity of the entropy functional \cite{Lieb2005}, and hence fulfills an area law whenever its quantum analog does. 

Exemplary, we consider a particle in a finite volume $L$ whose energy is sufficiently large compared to the infrared cutoffs set by the two subregions sizes, \textit{i.e.}, $\omega (p) \gg 1/l, 1/(L-l)$ \footnote{Otherwise, a quasi-particle state is locally indistinguishable from the ground state}. This condition is fulfilled for large momenta, a regime which has been analyzed in \cite{CastroAlvaredo2018a,CastroAlvaredo2018b,CastroAlvaredo2019a,CastroAlvaredo2019b} using the rather involved method of branch-point twist fields, but also for large masses. After straightforward analytic calculations, we find the scalings \cite{SM}
\begin{equation}
    \delta S_r [\mathcal{O}_{k,A}] = \frac{1}{1-r} \ln \left[ 1 + \sum_{i=1}^r a^{\mathcal{O}}_{r,i} \, \left( \frac{l}{L} \right)^i \right],
    \label{eq:SubtractedRenyiEntropiesParticle}
\end{equation}
for integer $r$ and real coefficients $a^{\mathcal{O}}_{r,i} < \infty$, see \autoref{fig:Entropies} \textbf{d)} and \textbf{e)} \footnote{For $r \notin \mathbb{N}  \setminus \{0,1 \}$, the sum over $i$ extends to infinity}. Since the polynomial inside the logarithm is bounded for $l/L \in [0,1]$, there exists a linear upper bound such that $\delta S_r [\mathcal{O}_{k,A}] \le [\ln (l/L)]/(1-r) +$const., which proves the area law for subtracted classical entropies of a particle. For small intervals $l \ll L$, we find relations reminiscent of the Gaussian case, namely $\delta S_r [\mathcal{W}_{k,A}] = 2 l/L = \delta S_2 [\boldsymbol{\rho}_{k,A}]$ and $\delta S_r [\mathcal{O}_{k,A}] = l/L = \delta S_2 [\boldsymbol{\rho}_A]/2$ otherwise, to leading order in $l/L$ [see linear inclines up to $l/L \lesssim 0.15$ in \autoref{fig:Entropies} \textbf{d)} and \textbf{e})].

The simplicity of \eqref{eq:SubtractedRenyiEntropiesParticle} enables the classification of the entanglement surplus of a particle via \eqref{eq:SubtractedClassicalRenyi2WignerEntropyAndMutualInformationGeneralRelation}. We find $\delta S_2 [\mathcal{W}_{k,A}] = - \ln \left[ (l/L)^2 + (1-l/L)^2 \right]$, which is in agreement with but goes beyond the validity of the formulae in \cite{CastroAlvaredo2018a,CastroAlvaredo2018b,CastroAlvaredo2019a,CastroAlvaredo2019b}. Interestingly, this result [black curves in \autoref{fig:Entropies} \textbf{d)} and \textbf{e)}] corresponds to the Rényi-Shannon-2 entropy of a coin with probability $l/L$. Heuristically, for large energies, the particle is delocalized over the full interval $[0,L]$, and thus the probability of the particle lying in the subregion $[0,l]$ is precisely $l/L$. While \eqref{eq:SubtractedRenyiEntropiesParticle} still holds for slow but heavy particles -- compare the black curve in \autoref{fig:Entropies} \textbf{f)} with the points representing numerical results for $p=0, m=10$ -- the entropy becomes largely independent of the subsystem size for smaller masses $m \lesssim 1/L$ (see transition from black to orange curves), since the particle is then confined to a small region.  

\textit{Discussion} --- We have shown that suitably chosen classical entropies and mutual informations exhibit area law-like scaling -- just as their quantum analogs. Being substantially simpler to evaluate theoretically \textit{and} to estimate from sparse data in an experimental setting (reliable estimations of such entropies were found for $10^4$ samples, see \cite{PRL,PRA}), we consider classical information measures inevitable tools to probe quantum phenomena, especially when state tomography is infeasible. We emphasize the experimental accessibility of the marginals \cite{Braunstein1990,Leonhardt1995,Welsch1999,Gross2011,Walborn2011,Kunkel2018,Schneeloch2019} and the Husimi $Q$-distribution \cite{Noh1991,Noh1992,Leibfried1996,Kirchmair2013,Haas2014,Strobel2014,Barontini2015,Wang2016,Gaerttner2017,Landon2018,Kunkel2019,Kunkel2021}, together with the generality of our approach, as all distributions can be defined analogously for fermionic and spin systems, see \cite{Zhang1990}. Thus, classical entropies pave the way for \textit{accessing} a variety of phenomena, including, \textit{e.g.}, local thermalization \cite{Kaufman2016,PRL}, quantum phase transitions \cite{Osborne2002}, topological order \cite{Kitaev2006}, and information scrambling \cite{Xu2022}.

\textit{Acknowledgements} --- I thank my colleagues Yannick Deller, Martin Gärttner, Markus K. Oberthaler, Moritz Reh, and Helmut Strobel for various insightful discussions and their valuable comments on earlier versions of the manuscript. I acknowledge support from the European Union under project ShoQC within the ERA-NET Cofund in Quantum Technologies (QuantERA) program, as well as from the F.R.S.- FNRS under project CHEQS within the Excellence of Science (EOS) program.


\bibliography{references.bib}

\end{document}


\title{Supplementary material}

\maketitle

The supplementary material is structured as follows. Details on the lattice regularization as well as the relativistic scalar field are provided in \autoref{sec:ScalarField}, which includes the wave functionals of all states of our interest, see \autoref{subsubsec:WaveFunctionals} (experienced readers may want to omit these sections). Then, we compute all phase space distributions in \autoref{sec:PhaseSpaceDistributions}. Here, the global distributions follow immediately from the wave functionals, see \autoref{subsec:GlobalDistributions}, while for the local distributions (\autoref{subsec:LocalDistributions}) we first derive some useful formulae for Gaussian integrals, see \autoref{subsubsec:GaussianIntegralsI}. Using similar formulae (\autoref{subsubsec:GaussianIntegralsII}), we thereupon calculate analytic formulae for the subtracted Rényi entropies in \autoref{sec:SubtractedEntropies}. A rigorous proof for the area law of classical mutual informations of a thermal state is provided in \autoref{sec:AreaLawsMutualInformations}. Supplementary figures are presented in \autoref{sec:Figures}, including an extended analysis of the central charge (\autoref{subsec:CentralCharge}), the correlations being present in the thermal state (\autoref{subsec:CorrelationsThermalState}), and the higher-order subtracted Rényi entropies of the marginal and Husimi $Q$-distributions (\autoref{subsec:HigherOrderRenyiEntropies}).

\section{Relativistic scalar field on the lattice}
\label{sec:ScalarField}
To highlight the roles of continuum and infinite volume limits for the various quantities of our interest, we work with a lattice description throughout the supplementary manuscript. Here, we give details on the regularization procedure as well as the discretized scalar field theory. 

\subsection{Finite lattice}
\label{subsec:FiniteLattice}
\begin{figure}[h!]
    \centering
    \includegraphics[width=0.33\columnwidth]{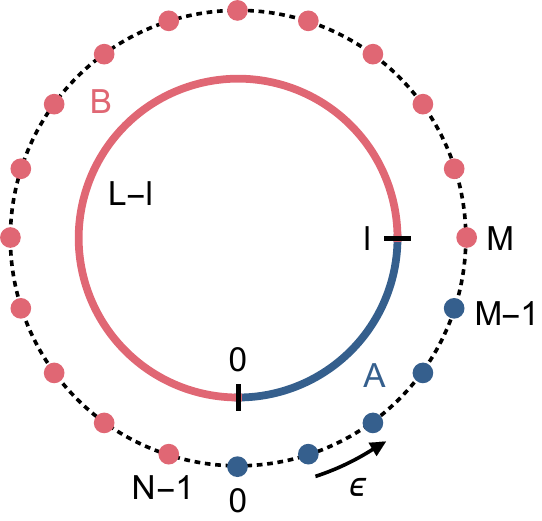}
    \caption{Illustration of the regularization procedure. Spatial positions $x$ are discretized to $N$ lattice points with separation $\epsilon$ and restricted to a finite interval of length $L=N \epsilon$ (circles). This procedure is applied to subregion $A$ (blue) and its complement $B$ (red) as well.}
    \label{fig:SpatialSubregion}
\end{figure}

We restrict the positions $x$ to a finite interval $x \in [0, L]$ with periodic boundary conditions, i.e. $\phi (0) = \phi(L)$, such that the geometry of the problem becomes isomorphic to the circle $S^1$ (see \autoref{fig:SpatialSubregion}). Also, we discretize the spatial positions according to
\begin{equation}
    x \to \epsilon \, j,
    \label{eq:PositionsDiscrete}
\end{equation}
with the discrete label $j \in \mathfrak{J} = \{0, ..., N-1\}$ and $\epsilon > 0$ being the lattice spacing, such that $L = N \epsilon$. In this scenario, the momenta are discrete as well
\begin{equation}
    p \to \eta \, k,
    \label{eq:MomentaDiscrete}
\end{equation}
with $\eta = (2 \pi)/L > 0$ being the spacing in momentum space and where the momentum label $k \in \mathfrak{K}$ depends on whether $N$ is even or odd, i.e.
\begin{equation}
    \mathfrak{K} = \begin{cases}
        \{- \frac{N-1}{2}, \dots, \frac{N-1}{2} \} & \text{ for } N \text{ odd}, \\
        \{- \frac{N}{2} + 1, \dots, \frac{N}{2} \} & \text{ for } N \text{ even}.
    \end{cases}
    \label{eq:MomentaDiscreteSet}
\end{equation}
Similarly, for the subinterval $A$ (blue arc and points), the discretization implies that the corresponding spatial indices are drawn from $\mathfrak{J}_A = \{0, ..., M-1 \}$, such that $A$ contains $M$ lattice points with $l=M\epsilon$. The indices $j \in \mathfrak{J}_B$ corresponding to the complement $B$ (red arc and points) run from $M$ to $N-1$, which is a total of $N-M$ lattice points, where now $L-l = (N-M)\epsilon$.

In this description, the continuum limit means taking the limits $\epsilon \to 0$ and $N,M \to \infty$ while keeping $L = N \epsilon$ and $l = M \epsilon$ fixed. Similarly, the infinite volume limit is obtained for $L,l \to \infty$ and $N \to \infty$ with $\epsilon = L/N = l/M$ fixed. In this sense, $1/\epsilon$ provides an ultraviolet cutoff scale, while $1/L$ acts as an infrared cutoff.

\subsection{Correspondence rules}
\label{subsec:CorrespondenceRules}
We write for the fields in position space
\begin{equation}
    \phi (x) \to \phi_j \equiv \phi (\epsilon j), \quad \pi (x) \to \pi_j \equiv \pi (\epsilon j),
    \label{eq:FieldsDiscretePositionSpace}
\end{equation}
while in momentum space we denote the fields by
\begin{equation}
    \tilde{\phi} (p) \to \tilde{\phi}_k \equiv \tilde{\phi} (\eta k), \quad \tilde{\pi} (p) \to \tilde{\pi}_k \equiv \tilde{\pi} (\eta k).
    \label{eq:FieldsDiscreteMomentumSpace}
\end{equation}
The regularization procedure implies a few replacement rules for typical terms appearing in Hamiltonians and functional distributions. We will employ the position-space identities
\begin{equation}
    \int \mathrm{d}x \to \sum_{j \in \mathfrak{J}} \epsilon, \quad \partial_x \phi (x) \to \frac{\phi_{j+1} - \phi_j}{\epsilon}, \quad \delta(x-x') \to \frac{\delta_{jj'}}{\epsilon},
    \label{eq:DiscretizationRulesPositionSpace}
\end{equation}
while in momentum space we will use
\begin{equation}
    \int \frac{\mathrm{d}p}{2 \pi} \to \sum_{k \in \mathfrak{K}} \frac{\eta}{2 \pi}, \quad \delta(p-p') \to \frac{\delta_{k k'}}{\eta}.
    \label{eq:DiscretizationRulesMomentumSpace}
\end{equation}
Momentum-space expressions and position-space expressions are related via a discrete Fourier transformation and its inverse, i.e.
\begin{equation}
    \tilde{\phi} (p) = \int \mathrm{d}x \, e^{- i x p} \phi (x) \to \tilde{\phi}_k = \sum_{j \in \mathfrak{J}} \epsilon \, e^{-i \epsilon j \eta k} \phi_j, \quad \phi (x) = \int \frac{\mathrm{d}p}{2\pi} \, e^{i x p} \tilde{\phi} (p) \to \phi_j = \sum_{k \in \mathfrak{K}} \frac{\eta}{2 \pi} \, e^{i \epsilon j \eta k} \tilde{\phi}_k,
    \label{eq:DiscreteFourierTransform}
\end{equation}
respectively, yielding the discretized integral representations of the Kronecker-$\delta$'s
\begin{equation}
    \begin{split}
        \delta (x-x') &= \int \frac{\mathrm{d}p}{2 \pi} \, e^{i (x-x')p} \to \frac{\delta_{j j'}}{\epsilon} = \sum_{k \in \mathfrak{K}} \frac{\eta}{2 \pi} \, e^{i \epsilon (j - j') \eta k}, \\ 
        2 \pi \delta (p-p') &= \int \mathrm{d}x \, e^{ix (p-p')} \to 2 \pi \frac{\delta_{k k'}}{\eta} = \sum_{j \in \mathfrak{J}} \epsilon \, e^{i \epsilon j \eta (k-k')}.
    \end{split}
    \label{eq:DiscreteKroneckerDeltas}
\end{equation}
Further, functional derivatives are replaced by ordinary derivatives
\begin{equation}
    \delta_{\phi (x')} \phi (x) = \delta (x - x') \to \partial_{\phi_{j'}} \phi_j = \frac{\delta_{j j'}}{\epsilon}, \quad  \delta_{\tilde{\phi} (p')} \tilde{\phi} (p) = 2 \pi \delta (p - p') \to \partial_{\tilde{\phi}_{k'}} \tilde{\phi}_k = 2 \pi \frac{\delta_{k k'}}{\eta}.
    \label{eq:DiscreteFunctionalDerivatives}
\end{equation}
The defining equation for the inverse of a matrix in position space is
\begin{equation}
    \int \mathrm{d} x' \, \mathcal{M} (x,x') \mathcal{M}^{-1} (x',x'') = \delta (x-x'') \, \to \, \sum_{j' \in \mathfrak{J}} \epsilon \, \mathcal{M}_{j j'} \mathcal{M}_{j' j''}^{-1} = \frac{1}{\epsilon} \delta_{j j''},
    \label{eq:DiscreteInverseMatrixPositionSpace}
\end{equation}
and analogously for momentum space, \textit{i.e.},
\begin{equation}
    \int \frac{\mathrm{d} p'}{2 \pi} \, \mathcal{N} (p,p') \mathcal{N}^{-1} (p',p'') = 2 \pi \delta (p-p'') \, \to \, \sum_{k' \in \mathfrak{K}} \frac{\eta}{2 \pi} \, \mathcal{N}_{k k'} \mathcal{N}_{k' k''}^{-1} = 2 \pi \frac{\delta_{k k''}}{\eta}.
    \label{eq:DiscreteInverseMatrixMomentumSpace}
\end{equation}
Similarly, traces and determinants acquire additional factors. More precisely, for diagonal matrices $\mathcal{M}$ and $\mathcal{N}$, we have
\begin{equation}
    \Tr \{ \mathcal{M} \} = \int \mathrm{d}x \, \mathcal{M} (x,x) \to \sum_{j \in \mathfrak{J}} \epsilon \, \mathcal{M}_{j j}, \quad \det ( \mathcal{M}) = \prod_{x \in \mathbb{R}} \mathcal{M} (x,x) \to \prod_{j \in \mathfrak{J}} \epsilon \, \mathcal{M}_{j j}.
    \label{eq:DiscreteDetTrPositionSpace}
\end{equation}
The momentum-space expressions follow from applying the inverse Fourier transformation \eqref{eq:DiscreteFourierTransform}, which yields
\begin{equation}
    \Tr \{ \mathcal{N} \} = \int \frac{\mathrm{d}p}{2 \pi} \, \mathcal{N} (p,p) \to \sum_{k \in \mathfrak{K}} \frac{\eta}{2 \pi} \, \mathcal{N}_{k k}, \quad \det ( \mathcal{N}) = \prod_{p \in \mathbb{R}} \mathcal{N} (p,p) \to \prod_{k \in \mathfrak{K}} \frac{\eta}{2 \pi} \, \mathcal{N}_{k k}.
    \label{eq:DiscreteDetTrMomentumSpace}
\end{equation}
At last, we define the functional integral measures over field configurations as
\begin{equation}
    \mathcal{D} \phi (x) = \prod_{j \in \mathfrak{J}} \mathrm{d} \phi_j \sqrt{\epsilon}, \quad \mathcal{D} \phi (p) = \prod_{k \in \mathfrak{K}} \mathrm{d} \phi_k \sqrt{\frac{\eta}{2 \pi}},
    \label{eq:FunctionalIntegralMeasures}
\end{equation}
and analogously for $\pi$. Let us note here that the normalizations of both measures are related to the normalization constants appearing in the functional distributions for the ground state required to ensure $\braket{0 | 0} = 1$ and also that all classical distributions are naturally normalized to unity. More precisely, with the normalization given in Eq. \eqref{eq:FunctionalIntegralMeasures}, the functional normalization constants become infinite products without any additional term $\sim \delta (0)$ in the continuum as well as in the infinite volume limit, see \autoref{subsubsec:WaveFunctionals}.

\subsection{Lattice theory}
\label{subsec:Setup}

\subsubsection{Hamiltonian}
We discretize the Hamiltonian given in Eq. (\textcolor{blue}{1}) in the main text using the replacement rules discussed in \autoref{subsec:CorrespondenceRules}, which leads to
\begin{equation}
   \boldsymbol{H} = \frac{1}{2} \sum_{j \in \mathfrak{J}} \epsilon \left[ \boldsymbol{\pi}_j^2 + \frac{\left(\boldsymbol{\phi}_{j+1} - \boldsymbol{\phi}_j \right)^2}{\epsilon^2} + m^2 \boldsymbol{\phi}_j^2 \right],
    \label{eq:DiscreteHamiltonian}
\end{equation}
with the discretized commutation relations
\begin{equation}
    [\boldsymbol{\phi}_j, \boldsymbol{\pi}_{j'}] = i \frac{\delta_{j j'}}{\epsilon}
    \label{eq:DiscreteCCRsPositionSpace}
\end{equation}
understood. It is decoupled using the inverse discrete Fourier transformation given in \eqref{eq:DiscreteFourierTransform}, resulting in 
\begin{equation}
    \boldsymbol{H} = \frac{1}{2} \sum_{k \in \mathfrak{K}} \frac{\eta}{2 \pi} \, \left( \tilde{\boldsymbol{\pi}}_k \tilde{\boldsymbol{\pi}}_{-k} + \omega_k^2 \, \tilde{\boldsymbol{\phi}}_k \tilde{\boldsymbol{\phi}}_{-k} \right).
    \label{eq:DiscreteHamiltonianDiagonal}
\end{equation}
In momentum space, the fields are complex-valued with commutation relations
\begin{equation}
    [\tilde{\boldsymbol{\phi}}_k, \tilde{\boldsymbol{\pi}}_{k'}] = i \, 2 \pi \frac{\delta_{k -k'}}{\eta},
    \label{eq:DiscreteCCRsMomentumSpace}
\end{equation}
and are constrained by $\tilde{\boldsymbol{\phi}}_{-k} = \tilde{\boldsymbol{\phi}}_k^*$ and $\tilde{\boldsymbol{\pi}}_{-k} = \tilde{\boldsymbol{\pi}}_k^*$ since the position-space fields are real. Further, the relativistic dispersion relation $\omega^2 (p) = m^2 + p^2$ has the discrete analog
\begin{equation}
    \omega_k^2 = m^2 + \frac{4}{\epsilon^2} \sin^2 \left( \frac{\epsilon \eta k}{2} \right).
    \label{eq:DiscreteDispersionRelation}
\end{equation}
Also, we introduce mode operators in momentum space
\begin{equation}
    \tilde{\boldsymbol{a}}_k = \frac{1}{\sqrt{2}} \left( \sqrt{\omega_k} \, \tilde{\boldsymbol{\phi}}_k + i \frac{1}{\sqrt{\omega_k}} \tilde{\boldsymbol{\pi}}_k \right), \quad \tilde{\boldsymbol{a}}^{\dagger}_k = \frac{1}{\sqrt{2}} \left( \sqrt{\omega_k}\, \tilde{\boldsymbol{\phi}}_{-k} - i \frac{1}{\sqrt{\omega_k}} \tilde{\boldsymbol{\pi}}_{-k} \right),
    \label{eq:DiscreteModeOperatorsMomentumSpace}
\end{equation}
with bosonic commutation relations
\begin{equation}
    [\tilde{\boldsymbol{a}}_k, \tilde{\boldsymbol{a}}^{\dagger}_{k'}] = 2 \pi \, \frac{\delta_{k k'}}{\eta} \mathds{1},
    \label{eq:DiscreteModeOperatorsCommutatorMomentumSpace}
\end{equation}
(note here that $[\boldsymbol{a}_j, \boldsymbol{a}_{j'}^{\dagger}] = (1/\epsilon) \delta_{j j'} \mathds{1}$ in position space), which diagonalize the Hamiltonian \eqref{eq:DiscreteHamiltonianDiagonal}, to wit
\begin{equation}
    \boldsymbol{H} = \sum_{k \in \mathfrak{K}} \frac{\eta}{2 \pi} \, \omega_k \left( \tilde{\boldsymbol{a}}_k^{\dagger} \tilde{\boldsymbol{a}}_k + \frac{1}{2} \frac{2 \pi}{\eta} \right).
    \label{eq:DiscreteHamiltonianDiagonalModeOperators}
\end{equation}
This form shows that the annihilation operator singles out the Hamiltonian's ground state via 
\begin{equation}
    \tilde{\boldsymbol{a}}_k \ket{0} = 0 \quad \forall k \in \mathfrak{K},
\end{equation}
with the ground state energy $E_0 = \braket{0 | \boldsymbol{H} | 0} = \sum_{k \in \mathfrak{K}} \omega_k/2$.

\subsubsection{Wave functionals}
\label{subsubsec:WaveFunctionals}
Next, we derive explicit expressions for all wave functionals of the states considered in the main text (the expressions for the momentum field wave functionals follow analogously). We start with computing the ground state wave functional $\Psi_0 [\phi] = \braket{\phi | 0}$, which is the solution to the stationary Schrödinger equation $\boldsymbol{H} \Psi_0 [\phi] = E_0 \Psi_0 [\phi]$. In momentum space, the latter becomes a standard second-order (integro-)differential equation
\begin{equation}
    \frac{1}{2} \sum_{k \in \mathfrak{K}} \frac{\eta}{2 \pi} \, \left( - \partial_{\tilde{\phi}_k} \partial_{\tilde{\phi}_{-k}} + \omega_k^2 \, \tilde{\phi}_k \tilde{\phi}_{-k} \right) \tilde{\Psi}_0 [\tilde{\phi}] = \sum_{k \in \mathfrak{K}} \frac{\omega_k}{2} \tilde{\Psi}_0 [\tilde{\phi}],
    \label{eq:SchrödingerEquation}
\end{equation}
whose solution is of Gaussian form
\begin{equation}
    \tilde{\Psi}_0 [\tilde{\phi}] = \frac{1}{\sqrt{\tilde{Z}^f_{0}}} \, e^{- \frac{1}{4} \sum_{k,k' \in \mathfrak{K}} (\frac{\eta}{2 \pi})^2 \, \tilde{\phi}_k (\tilde{\gamma}_0^f)_{j j'}^{-1} \tilde{\phi}_{-k}},
    \label{eq:GroundStateFunctionalsMomentumSpace}
\end{equation}
with covariance matrices and normalizations 
\begin{equation}
    (\tilde{\gamma}_0^{f})_{k k'} = \frac{2 \pi}{\eta} \frac{1}{2\omega_k} \delta_{k k'}, \quad (\tilde{\gamma}_0^{g})_{k k'} = \frac{2 \pi}{\eta} \frac{\omega_k}{2} \delta_{k k'}, \quad \tilde{Z}_0^f = \prod_{k \in \mathfrak{K}} \sqrt{\frac{\pi}{\omega_k}}, \quad \quad \tilde{Z}_0^g = \prod_{k \in \mathfrak{K}} \sqrt{\pi \omega_k},
    \label{eq:GroundStateMarginalCovarianceNormalizationMomentumSpace}
\end{equation}
for the field $\phi$ (index $f$) and the momentum field $\pi$ (index $g$),
respectively. The position-space expression follows after a discrete Fourier transformation \eqref{eq:DiscreteFourierTransform} of the fields 
\begin{equation}
    \Psi_0 [\phi] = \frac{1}{\sqrt{Z^f_{0}}} \, e^{- \frac{1}{4} \sum_{j,j' \in \mathfrak{J}} \epsilon^2 \, \phi_j (\gamma_0^f)^{-1} \phi_{j'}},
    \label{eq:GroundStateWaveFunctionalsPositionSpace}
\end{equation}
with the well-known two-point correlation functions
\begin{equation}
    (\gamma_0^f)_{j j'} = \sum_{k \in \mathfrak{K}} \frac{\eta}{2 \pi} \frac{1}{2 \omega_k} \cos \left[\epsilon (j-j') \eta k \right], \quad (\gamma^g_0)_{j j'} = \sum_{k \in \mathfrak{K}} \frac{\eta}{2 \pi} \, \frac{\omega_k}{2} \, \cos \left[\epsilon (j-j') \eta k \right],
    \label{eq:GroundStateMarginalCovarianceNormalizationPositionSpace}
\end{equation}
and unaltered normalizations $Z_0^f = \tilde{Z}_0^f, Z_0^g = \tilde{Z}_0^g$. Note here the useful relation $(\gamma_0^{f})^{-1}_{j j'} = 4 (\gamma_0^{g})_{j j'}$.

A quasi-particle excitation with momentum $k$ is generated by acting on the ground-state wave functional with the corresponding creation operator transformed to position space
\begin{equation}
    \Psi_k [\phi] = \sqrt{\frac{\eta}{2 \pi}} \tilde{\boldsymbol{a}}^{\dagger}_k \Psi_0 [\phi] = \sqrt{\frac{\eta}{2 \pi}} \left( \sum_{j \in \mathfrak{J}} \epsilon \, e^{- i \epsilon j \eta k} \boldsymbol{a}^{\dagger}_j \right) \Psi_0 [\phi] = \sum_{j \in \mathfrak{J}} \epsilon \, e^{- i \epsilon j \eta k} \frac{\phi_j}{\sqrt{(\tilde{\gamma}_0^f)_{k k}}} \, \Psi_0 [\phi],
\end{equation}
where we used \eqref{eq:DiscreteModeOperatorsMomentumSpace} in the Schrödinger representation and identified the diagonal elements of \eqref{eq:GroundStateMarginalCovarianceNormalizationMomentumSpace} as
\begin{equation}
    (\tilde{\gamma}_0^f)_{k k} = \frac{2 \pi}{\eta} \frac{1}{2 \omega_k}, \quad (\tilde{\gamma}_0^g)_{k k} = \frac{2 \pi}{\eta} \frac{\omega_k}{2}.
    \label{eq:GroundStateMarginalCovarianceDiagonalEntries}
\end{equation}

At last, the density matrix of the thermal state 
\begin{equation}
    \rho_T [\phi, \phi'] = \frac{1}{Z_T} \braket{\phi | e^{- \beta \boldsymbol{H}} | \phi'}
\end{equation}
follows after employing the Baker-Campbell-Hausdorff formula to separate fields and conjugate fields in the exponential, expanding the exponentials and using the overlap $\braket{\phi | \boldsymbol{\pi}^2_j | \phi'} = - i \partial^2_{\phi_j} \delta_{\phi \phi'}$. This results in
\begin{equation}
    \rho_T [\phi, \phi'] = \frac{1}{\sqrt{Z^f_{T}}} \, e^{- \frac{1}{2} \sum_{j,j' \in \mathfrak{J}} \epsilon^2 \, \phi_j (\gamma_T^f)^{-1}_{j j'} \phi'_{j'}},
\end{equation}
with the thermal correlators
\begin{equation}
    (\gamma_T^f)_{j j'} = \sum_{k \in \mathfrak{K}} \frac{\eta}{2 \pi} \frac{1}{2 \omega_k} \cos \left[\epsilon (j-j') \eta k \right] \left( 1 + 2 \braket{n_k}_{\text{BE}} \right), \quad (\gamma^g_T)_{j j'} = \sum_{k \in \mathfrak{K}} \frac{\eta}{2 \pi} \, \frac{\omega_k}{2} \, \cos \left[\epsilon (j-j') \eta k \right] \left( 1 + 2 \braket{n_k}_{\text{BE}} \right),
    \label{eq:ThermalStateMarginalCovariancePositionSpace}
\end{equation}
and Bose-Einstein statistics
\begin{equation}
    \braket{n_k}_{\text{BE}} = \frac{1}{e^{\beta \omega_k} - 1}.
\end{equation}

\subsubsection{Vacuum theory}
\label{subsubsec:VacuumTheory}
The vacuum in position space is defined as the ground state of a set of uncoupled oscillators. Starting from the lattice Hamiltonian \eqref{eq:DiscreteHamiltonian}, this implies that the derivative term coupling neighboring modes is absent in the uncoupled theory. Also, we have to respect the mass dimensions of the fundamental fields. In the relativistic field theory \eqref{eq:DiscreteHamiltonian}, the field has mass dimension $[\phi_j] = 0$, while the momentum field has mass dimension $[\pi_j] = 1$. Since the only available length scale for the uncoupled theory is the lattice spacing $\epsilon$, the mass in \eqref{eq:DiscreteHamiltonian} has to be replaced by the inverse lattice spacing to ensure correct mass dimensions, resulting in 
\begin{equation}
    \bar{\boldsymbol{H}} = \frac{1}{2} \sum_{j \in \mathfrak{J}} \epsilon \left[ \boldsymbol{\pi}_j^2 + \frac{1}{\epsilon^2} \boldsymbol{\phi}_j^2 \right].
    \label{eq:DiscreteHamiltonianVacuum}
\end{equation}
The Fourier transformation to momentum space becomes obsolete, as we already deal with a Hamiltonian of the form \eqref{eq:DiscreteHamiltonianDiagonal}, with the dispersion being the inverse lattice spacing
\begin{equation}
    \bar{\omega} = \frac{1}{\epsilon}.
    \label{eq:DiscreteDispersionRelationVacuum}
\end{equation}
This implies that the vacuum in position space is described by a wave functional of the form \eqref{eq:GroundStateMarginalCovarianceNormalizationPositionSpace} with diagonal covariances
\begin{equation}
    (\bar{\gamma}^f)_{j j'} = \frac{\epsilon}{2} \frac{\delta_{j j'}}{\epsilon}, \quad (\bar{\gamma}^g)_{j j'} = \frac{1}{2 \epsilon} \frac{\delta_{j j'}}{\epsilon}.
    \label{eq:VacuumStateMarginalCovarianceNormalizationPositionSpace}
\end{equation}

\section{Phase space distributions}
\label{sec:PhaseSpaceDistributions}
With the wave functionals at hand, we are ready to compute the global phase-space distributions. The local distributions then follow from integrating out the complementary degrees of freedom, for which an adapted version of the central identity of quantum field theory will be of great use.

\subsection{Global distributions}
\label{subsec:GlobalDistributions}

\subsubsection{Wigner $W$-distribution}
\label{subsubsec:WignerWDistributionsGlobal}
The ground state's Wigner $W$-distribution follows after a straightforward exercise in Gaussian integration
\begin{equation}
    \mathcal{W}_0 [\chi] = \int \frac{\mathcal{D} \phi'}{\pi} \Psi^*_0 [\phi - \phi'] \Psi_0 [\phi + \phi'] e^{2 i \sum_{j \in \mathfrak{J}} \epsilon \, \pi_j \phi'_j} = \frac{1}{Z_0^{\mathcal{W}}} e^{- \frac{1}{2} \sum_{j,j' \in \mathfrak{J}} \epsilon^2 \chi^T_j (\gamma_0^{\mathcal{W}})^{-1}_{j j'} \chi_{j'}}.
    \label{eq:GroundStateWignerWDistributionPositionSpace}
\end{equation}
Its covariance matrix and normalization decompose as
\begin{equation}
    \gamma^{\mathcal{W}}_0 = \gamma^{f}_0 \oplus \gamma_0^{g}, \quad Z_0^{\mathcal{W}} = \sqrt{\det \left( 2 \pi \gamma_0^{\mathcal{W}} \right)} = \prod_{k \in \mathfrak{K}} \pi = Z_0^f \, Z_0^g,
    \label{eq:GroundStateWignerWCovarianceNormalizationPositionSpace}
\end{equation}
revealing its product form $\mathcal{W}_0 [\chi] = f_0 [\phi] \times g_0 [\pi]$. For a quasi-particle, the situation is slightly more involved as higher-order Gaussian integrals appear. After a few steps, we find the expression
\begin{equation}
    \mathcal{W}_{k} [\chi] = \mathcal{W}_0 [\chi] \left\{ - 1 + \sum_{j,j' \in \mathfrak{J}} \epsilon^2 \, e^{-i \epsilon (j-j') \eta k} \left[ \frac{\phi_j \phi_{j'}}{(\tilde{\gamma}_0^f)_{kk}} + \frac{\pi_j \pi_{j'}}{(\tilde{\gamma}_0^g)_{kk}} - i \frac{2}{L} \left(\phi_j \pi_{j'} - \pi_j \phi_{j'} \right) \right] \right\},
    \label{eq:QuasiParticleWignerWDistributionPositionSpace}
\end{equation}
which is not of product form. For thermal equilibrium, we obtain the product distribution
\begin{equation}
    \mathcal{W}_T [\chi] = \frac{1}{Z_T^{\mathcal{W}}} e^{- \frac{1}{2} \sum_{j,j' \in \mathfrak{J}} \epsilon^2 \chi^T_j (\gamma_T^{\mathcal{W}})^{-1}_{j j'} \chi_{j'}},
    \label{eq:ThermalStateWignerWDistributionPositionSpace}
\end{equation}
with 
\begin{equation}
    \gamma^{\mathcal{W}}_T = \gamma^{f}_T \oplus \gamma_T^{g}, \quad Z_T^{\mathcal{W}} = \sqrt{\det \left(2 \pi \gamma_T^{\mathcal{W}} \right)} = Z_T^f \, Z_T^g.
    \label{eq:ThermalStateWignerWCovarianceNormalizationPositionSpace}
\end{equation} 

\subsubsection{Marginal distributions}
The marginal distribution of the vacuum over the field $\phi$ is computed using Born's rule
\begin{equation}
    f_0 [\phi] = \abs{\Psi_0 [\phi]}^2 = \frac{1}{Z_0^f} e^{- \frac{1}{2} \sum_{j,j' \in \mathfrak{J}} \epsilon^2 \phi_j (\gamma_0^{f})^{-1}_{j j'} \phi_{j'}},
    \label{eq:GroundStateMarginalDistributionPositionSpace}
\end{equation}
with the position covariance matrix \eqref{eq:GroundStateMarginalCovarianceNormalizationPositionSpace} understood and the normalization constant being related to the covariance matrix via $Z_0^{f} = \sqrt{\det (2 \pi \gamma_0^{f})}$. The quasi-particle expression is again obtained from Born's rule [or from marginalizing Eq. \eqref{eq:QuasiParticleWignerWDistributionPositionSpace}], to wit
\begin{equation}
    f_k [\phi] = \abs{\Psi_k [\phi]}^2 = f_0 [\phi] \, \sum_{j,j' \in \mathfrak{J}} \epsilon^2 \, e^{-i \epsilon (j-j') \eta k} \frac{\phi_j \phi_{j'}}{(\tilde{\gamma}^f_0)_{kk}},
    \label{eq:QuasiParticleMarginalDistributionPositionSpace}
\end{equation}
while in thermal equilibrium we have
\begin{equation}
    f_T [\phi] = \rho_T [\phi, \phi] = \frac{1}{Z_T^f} e^{- \frac{1}{2} \sum_{j,j' \in \mathfrak{J}} \epsilon^2 \phi_j (\gamma_T^{f})^{-1}_{j j'} \phi_{j'}},
    \label{eq:ThermalStateMarginalDistributionPositionSpace}
\end{equation}
where $Z_T^{f} = \sqrt{\det (2 \pi \gamma_T^{f})}$. The corresponding expressions for $g [\pi]$ are found analogously.

\subsubsection{Husimi $Q$-distribution}
By convolving Eq. \eqref{eq:GroundStateWignerWDistributionPositionSpace} with the vacuum, we find for the ground-state Husimi $Q$-distribution the expression
\begin{equation}
    \mathcal{Q}_0 [\chi] = \frac{1}{Z_0^{\mathcal{Q}}} e^{- \frac{1}{2} \sum_{j,j' \in \mathfrak{J}} \epsilon^2 \chi^T_j (\gamma^{\mathcal{Q}}_0)^{-1}_{j j'} \chi_{j'}},
    \label{eq:GroundStateHusimiQDistributionPositionSpace}
\end{equation}
with covariance matrix and normalization
\begin{equation}
    \gamma^{\mathcal{Q}}_0 = \gamma^{\mathcal{W}}_0 + \bar{\gamma}^{\mathcal{W}}, \quad Z_0^{\mathcal{Q}} = \sqrt{\det \left( 2 \pi \gamma_0^{\mathcal{Q}}\right)}.
    \label{eq:GroundStateHusimiQCovarianceNormalizationPositionSpace}
\end{equation}
Note here that the variances increase since
\begin{equation}
    (\gamma_0^{\mathcal{Q},f})_{j j'} = (\gamma^f_0)_{j j'} + \frac{1}{2} \delta_{j j'}, \quad (\gamma_0^{\mathcal{Q},g})_{j j'} = (\gamma^g_0)_{j j'} + \frac{1}{2 \epsilon^2} \delta_{j j'},
    \label{eq:HusimiQMarginalCovariance}
\end{equation}
showing that the marginal distributions of the Husimi $Q$-distribution are less localized than the true marginals. Similarly, a quasi-particle is described by
\begin{equation}
    \begin{split}
        \mathcal{Q}_{k} [\chi] &= \mathcal{Q}_0 [\chi] \sum_{j,j' \in \mathfrak{J}} \epsilon^2 \, e^{-i \epsilon (j-j') \eta k} \Bigg\{ \frac{(\tilde{\gamma}^f_0)_{kk}}{\left[(\tilde{\gamma}^f_0)_{kk} + \bar{\tilde{\gamma}}^{f}_{k k} \right]^2} \phi_j \phi_{j'} + \frac{(\tilde{\gamma}^g_0)_{kk}}{\left[(\tilde{\gamma}^g_0)_{kk} + \bar{\tilde{\gamma}}^{g}_{k k} \right]^2} \pi_j \pi_{j'} \\
        &\hspace{4.6cm}- i \frac{L}{2} \frac{1}{\left[(\tilde{\gamma}^f_0)_{kk} +\bar{\tilde{\gamma}}^{f}_{k k} \right]\left[(\tilde{\gamma}^g_0)_{kk} + \bar{\tilde{\gamma}}^{g}_{k k} \right]} \left(\phi_j \pi_{j'} - \pi_j \phi_{j'} \right) \Bigg\},
    \end{split}    \label{eq:QuasiParticleHusimiQDistributionPositionSpace}
\end{equation}
where 
\begin{equation}
    \bar{\tilde{\gamma}}^{f}_{k k} = \frac{2 \pi}{\eta}\frac{\epsilon}{2}, \quad \bar{\tilde{\gamma}}^{g}_{k k} = \frac{2 \pi}{\eta}\frac{1}{2\epsilon}.
\end{equation}
The prefactors can be simplified to
\begin{equation}
    \begin{split}
        \frac{(\tilde{\gamma}^f_0)_{kk}}{\left[(\tilde{\gamma}^f_0)_{kk} + \bar{\tilde{\gamma}}^{f}_{k k} \right]^2} &= \frac{2}{L} \frac{\omega_k}{(1 + \epsilon \omega_k)^2}, \\\frac{(\tilde{\gamma}^g_0)_{kk}}{\left[(\tilde{\gamma}^g_0)_{kk} + \bar{\tilde{\gamma}}^{g}_{k k} \right]^2} &= \frac{2}{L} \frac{\epsilon^2 \omega_k}{(1 + \epsilon \omega_k)^2}, \\
        \frac{1}{\left[(\tilde{\gamma}^f_0)_{kk} +\bar{\tilde{\gamma}}^{f}_{k k} \right]\left[(\tilde{\gamma}^g_0)_{kk} + \bar{\tilde{\gamma}}^{g}_{k k} \right]} &= \frac{4}{L^2} \frac{\epsilon \omega_k}{(1 + \epsilon \omega_k)^2}
    \end{split}
    \label{eq:QuasiParticleHusimiQPrefactors}
\end{equation}
Finally, the thermal Husimi $Q$-distribution reads
\begin{equation}
    \mathcal{Q}_T [\chi] = \frac{1}{Z_T^{\mathcal{Q}}} e^{- \frac{1}{2} \sum_{j,j' \in \mathfrak{J}} \epsilon^2 \chi^T_j (\gamma^{\mathcal{Q}}_T)^{-1}_{j j'} \chi_{j'}},
    \label{eq:ThermalStateHusimiQDistributionPositionSpace}
\end{equation}
with
\begin{equation}
    \gamma_T^{\mathcal{Q}} = \gamma_T^{\mathcal{W}} + \bar{\gamma}^{\mathcal{W}}, \quad Z_T^{\mathcal{Q}} = \sqrt{\det \left(2 \pi \gamma_T^{\mathcal{Q}} \right)}.
    \label{eq:ThermalStateHusimiQCovarianceNormalizationPositionSpace}
\end{equation}

\subsection{Local distributions}
\label{subsec:LocalDistributions}

\subsubsection{Intermezzo on Gaussian integrals}
\label{subsubsec:GaussianIntegralsI}
Ultimately, we are interested in the local distributions corresponding to subsystem $A$ by integrating out all degrees of freedom corresponding to subsystem $B$. We first note that all distributions of our interest can be decomposed as
\begin{equation}
    \mathcal{O} [\nu] = G^{\mathcal{O}} [\nu] \, \kappa^{\mathcal{O}} [\nu], 
    \label{eq:GaussianIntegralsGeneralDistribution}
\end{equation}
where $G^{\mathcal{O}} [\nu]$ is a Gaussian distribution 
\begin{equation}
    G^{\mathcal{O}} [\nu] = \frac{1}{Z^{\mathcal{O}}} \, e^{- \frac{1}{2} \nu^T (\gamma^{\mathcal{O}})^{-1} \nu}
    \label{eq:GaussianIntegralsGeneralGaussian}
\end{equation}
with zero mean, a covariance matrix $\gamma^{\mathcal{O}}$ of Toeplitz-form, and normalization $Z^{\mathcal{O}} = \sqrt{\det \left( 2 \pi \gamma^{\mathcal{O}} \right)}$ ensuring $\int \mathcal{D} \nu \, G^{\mathcal{O}} [\nu] = 1$, over the $u$-dimensional random vector $\nu = (\nu_0, ..., \nu_{u-1})^T$. Further, we assume $\kappa^{\mathcal{O}} [\nu]$ to be a real-valued quadratic form over $\nu$ up to a constant, \textit{i.e.},
\begin{equation}
    \kappa^{\mathcal{O}} [\nu] = \theta^{\mathcal{O}} + \nu^T \Theta^{\mathcal{O}} \nu,
\end{equation}
for some real constant $\theta^{\mathcal{O}}$ and a hermitian matrix $\Theta^{\mathcal{O}}$, which, in our case, corresponds to the matrix associated with the discrete Fourier transform \eqref{eq:DiscreteFourierTransform} up to prefactors. Then, a local distribution over subsystem $A$ is obtained by solving the integral
\begin{equation}
    \mathcal{O}_A [\nu_A] = \int \mathcal{D} \nu_B \, G^{\mathcal{O}} [\nu_A, \nu_B] \, \kappa^{\mathcal{O}} [\nu_A, \nu_B].
    \label{eq:GaussianIntegralsGeneralLocalDistribution1}
\end{equation}
We group the field $\nu$ in a two-dimensional vector with respect to the two spatial subsystems $A$ and $B$ of dimensions $v$ and $u-v$, respectively, such that
\begin{equation}
    \nu = (\nu_A, \nu_B)^T,
\end{equation}
with
\begin{equation}
    \nu_A = (\nu_0, ..., \nu_{v-1})^T, \quad \nu_B = (\nu_v, ..., \nu_{u-1})^T.
\end{equation}
We apply the same decomposition to the covariance matrix and the quadratic form
\begin{equation}
    \gamma^{\mathcal{O}} = \begin{pmatrix}
        \gamma^{\mathcal{O}}_A & \gamma^{\mathcal{O}}_M \\
        (\gamma^{\mathcal{O}}_M)^T & \gamma^{\mathcal{O}}_B
    \end{pmatrix}, \quad \Theta^{\mathcal{O}} = \begin{pmatrix}
        \Theta_A^{\mathcal{O}} & \Theta_M^{\mathcal{O}} \\
        (\Theta_M^{\mathcal{O}})^{\dagger} & \Theta_B^{\mathcal{O}}
    \end{pmatrix}.
\end{equation}
Since $\gamma^{\mathcal{O}}$ is a Toeplitz matrix, all its diagonal blocks are Toeplitz matrices as well. However, only the diagonal blocks $\gamma_A^{\mathcal{O}}$ and $\gamma_B^{\mathcal{O}}$ are symmetric in general, while the off-diagonal block $\gamma^{\mathcal{O}}_M$ describing correlations between $A$ and $B$ is not (its symmetry is restored only if subsystem $A$ spans precisely half of the total system, \textit{i.e.}, if $l=L/2$). Next, we define the covariance matrix' inverse, the so-called precision matrix, as
\begin{equation}
    (\gamma^{\mathcal{O}})^{-1} \equiv \Gamma^{\mathcal{O}} = 
    \begin{pmatrix}
        \Gamma^{\mathcal{O}}_A & \Gamma^{\mathcal{O}}_M \\
        (\Gamma^{\mathcal{O}}_M)^T & \Gamma^{\mathcal{O}}_B
    \end{pmatrix}.
\end{equation}
Since all six submatrices of the covariance matrix and its inverse are invertable, their components fulfill various useful relations, of which we recall the following three for later purposes
\begin{equation}
    (\gamma^{\mathcal{O}}_{A})^{-1} = \Gamma^{\mathcal{O}}_{A} - \Gamma^{\mathcal{O}}_{M}(\Gamma^G_{B})^{-1} (\Gamma^{\mathcal{O}}_{M})^T, \quad (\Gamma^G_{B})^{-1} = \gamma_{B}^{\mathcal{O}} - (\gamma_{M}^{\mathcal{O}})^T (\gamma_{A}^{\mathcal{O}})^{-1} \gamma_{M}^{\mathcal{O}}, \quad (\gamma^{\mathcal{O}}_{M})^T (\gamma_{A}^{\mathcal{O}})^{-1} = - (\Gamma^{\mathcal{O}}_{B})^{-1} (\Gamma^{\mathcal{O}}_{M})^T.
    \label{eq:GaussianIntegralsLocalCovarianceMatricesRelations}
\end{equation}
The first two are nothing but the formulae for the Schur complements of the blocks $\Gamma^{\mathcal{O}}_B$ and $\gamma_A^{\mathcal{O}}$, respectively, while the second is the matrix of the so-called regression coefficients. Now, performing the decomposition into $A$ and $B$ for the two elementary building blocks of the distributions of our interest yields
\begin{equation}
    \begin{split}
        G^{\mathcal{O}} [\nu_A, \nu_B] &= \frac{1}{Z^{\mathcal{O}}} e^{- \frac{1}{2} \left( \nu_A^T \Gamma_A^{\mathcal{O}} \nu_A + \nu_A^T \Gamma_M^{\mathcal{O}} \nu_B + \nu_B^T (\Gamma_M^{\mathcal{O}})^T \nu_A + \nu_B^T \Gamma^{\mathcal{O}}_B \nu_B \right)}, \\ \kappa^{\mathcal{O}} [\nu_A, \nu_B] &= \theta^{\mathcal{O}} + \nu_A^T \Theta_A^{\mathcal{O}} \nu_A + \nu_A^T \Theta_M^{\mathcal{O}} \nu_B + \nu_B^T (\Theta_M^{\mathcal{O}})^{\dagger} \nu_A + \nu_B^T \Theta_B^{\mathcal{O}} \nu_B.
    \end{split}
    \label{eq:GaussianIntegralsGeneralDistributionDecomposition}
\end{equation}
Then, introducing a source term $\zeta$ for subsystem $B$ and using the latter relations, the integral \eqref{eq:GaussianIntegralsGeneralLocalDistribution1} can be rewritten as 
\begin{equation}
    \begin{split}
        \mathcal{O}_A [\nu_A] &= \frac{1}{Z^{\mathcal{O}}} \, e^{- \frac{1}{2} \nu_A^T \Gamma^{\mathcal{O}}_A \nu_A} \int \mathcal{D} \nu_B \, e^{- \frac{1}{2} \nu_B^T \Gamma^{\mathcal{O}}_B \nu_B - \nu_A^T \Gamma^{\mathcal{O}}_M \nu_B + \zeta^T \nu_B} \kappa^{\mathcal{O}} [\nu_A, \nu_B] \bigg \vert_{\zeta = 0} \\
        &= \frac{1}{Z^{\mathcal{O}}} \, e^{- \frac{1}{2} \nu_A^T \Gamma^{\mathcal{O}}_A \nu_A} \kappa^{\mathcal{O}} [\nu_A, \partial_\zeta] \int \mathcal{D} \nu_B \, e^{ - \frac{1}{2} \nu_B^T \Gamma^{\mathcal{O}}_B \nu_B + \left( \zeta^T - \nu_A^T \Gamma^{\mathcal{O}}_M \right) \nu_B} \bigg \vert_{\zeta = 0},
    \end{split}
    \label{eq:GaussianIntegralsGeneralLocalDistribution2}
\end{equation}
where we performed the well-known derivative trick in the second step to pull the function $\kappa^{\mathcal{O}} [\nu_A, \nu_B]$ out of the integral. The remaining integral is a standard Gaussian integral with an additional linear term, which, together with \eqref{eq:GaussianIntegralsLocalCovarianceMatricesRelations}, evaluates to
\begin{equation}
    \begin{split}
        \mathcal{O}_A [\nu_A] &= \frac{\sqrt{\det \left[ 2 \pi (\Gamma^{\mathcal{O}}_B)^{-1} \right]}}{Z^{\mathcal{O}}} \, e^{- \frac{1}{2} \nu_A^T \left[ \Gamma^{\mathcal{O}}_A - \Gamma^{\mathcal{O}}_{M}(\Gamma^{\mathcal{O}}_{B})^{-1} (\Gamma^{\mathcal{O}}_{M})^T \right] \nu_A} \kappa^{\mathcal{O}} [\nu_A, \partial_\zeta] \, e^{\frac{1}{2} \zeta^T (\Gamma^{\mathcal{O}}_B)^{-1} \zeta - \zeta^T (\Gamma^{\mathcal{O}}_B)^{-1} (\Gamma^{\mathcal{O}}_M)^T \nu_A} \bigg \vert_{\zeta = 0} \\
        &= \frac{\sqrt{\det \left[ 2 \pi (\Gamma^{\mathcal{O}}_B)^{-1} \right]}}{Z^{\mathcal{O}}} \, e^{- \frac{1}{2} \nu_A^T (\gamma^{\mathcal{O}}_A)^{-1} \nu_A} \kappa^{\mathcal{O}} [\nu_A, \partial_{\zeta}] \, e^{\frac{1}{2} \zeta^T (\Gamma^{\mathcal{O}}_B)^{-1} \zeta + \zeta^T (\gamma^{\mathcal{O}}_M)^{T} (\gamma^{\mathcal{O}}_A)^{-1} \nu_A} \bigg \vert_{\zeta = 0}.
    \end{split}
\end{equation}
For brevity, we define
\begin{equation}
    G_A [\nu_A] \equiv \frac{1}{Z^{\mathcal{O}}_A} \, e^{- \frac{1}{2} \nu_A^T (\gamma^{\mathcal{O}}_A)^{-1} \nu_A}, \quad Z^{\mathcal{O}}_A \equiv \frac{Z^{\mathcal{O}}}{\sqrt{\det \left[ 2 \pi (\Gamma^{\mathcal{O}}_B)^{-1} \right]}} = \sqrt{\det \left( 2 \pi \gamma^{\mathcal{O}}_A \right)}, \quad \Omega^{\mathcal{O}} \equiv (\gamma^{\mathcal{O}}_M)^T (\gamma^{\mathcal{O}}_A)^{-1},
\end{equation}
such that
\begin{equation}
    \mathcal{O}_A [\nu_A] = G^{\mathcal{O}}_A [\nu_A] \, \kappa^{\mathcal{O}} [\nu_A, \partial_{\zeta}] \, e^{\frac{1}{2} \zeta^T (\Gamma^{\mathcal{O}}_B)^{-1} \zeta + \zeta^T \Omega^{\mathcal{O}} \nu_A} \bigg \vert_{\zeta = 0}.
\end{equation}
The remaining derivatives can be computed using \eqref{eq:GaussianIntegralsGeneralDistributionDecomposition}, resulting in
\begin{equation}
    \begin{split}
        &\kappa^{\mathcal{O}} [\nu_A, \partial_{\zeta}] \, e^{\frac{1}{2} \zeta^T (\Gamma^{\mathcal{O}}_B)^{-1} \zeta + \zeta^T \Omega^{\mathcal{O}} \nu_A} \bigg \vert_{\zeta = 0} \\
        &= \theta^{\mathcal{O}} + \nu_A^T \Theta_A^{\mathcal{O}} \nu_A + \left[ \nu_A^T \Theta^{\mathcal{O}}_M \partial_{\zeta} + \partial^T_{\zeta} (\Theta^{\mathcal{O}}_M)^{\dagger} \nu_A + \partial_{\zeta}^T \Theta^{\mathcal{O}}_B \partial_{\zeta} \right] e^{\frac{1}{2} \zeta^T (\Gamma^{\mathcal{O}}_B)^{-1} \zeta + \zeta^T \Omega^{\mathcal{O}} \nu_A} \bigg \vert_{\zeta = 0} \\
        &= \theta^{\mathcal{O}} + \nu_A^T \Theta^{\mathcal{O}}_A \nu_A + \nu_A^T \Theta^{\mathcal{O}}_M \Omega^{\mathcal{O}} \nu_A + \nu_A^T (\Omega^{\mathcal{O}})^T (\Theta^{\mathcal{O}}_M)^{\dagger} \nu_A + \Tr_B \{\Theta^{\mathcal{O}}_B (\Gamma^{\mathcal{O}}_B)^{-1} \} + \nu_A^T (\Omega^{\mathcal{O}})^T \Theta^{\mathcal{O}}_B \Omega^{\mathcal{O}} \nu_A \\
        &= \theta^{\mathcal{O}} + \Tr_B \{\Theta^{\mathcal{O}}_B (\Gamma^{\mathcal{O}}_B)^{-1} \} + \nu_A^T \left[\Theta_A^{\mathcal{O}} + \Theta^{\mathcal{O}}_M \Omega^{\mathcal{O}} + (\Omega^{\mathcal{O}})^T (\Theta^{\mathcal{O}}_M)^{\dagger} + (\Omega^{\mathcal{O}})^T \Theta_B^{\mathcal{O}} \Omega^{\mathcal{O}} \right] \nu_A.
    \end{split} 
\end{equation}
To state the final formula in a concise form, we introduce a real-valued number and a Hermitian matrix,
\begin{equation}
    \lambda_A^{\mathcal{O}} \equiv \theta^{\mathcal{O}} + \Tr_B \{\Theta^{\mathcal{O}}_B (\Gamma^{\mathcal{O}}_B)^{-1} \}, \quad \Lambda_A^{\mathcal{O}} \equiv \Theta^{\mathcal{O}}_A + \Theta^{\mathcal{O}}_M \Omega^{\mathcal{O}} + (\Omega^{\mathcal{O}})^T (\Theta^{\mathcal{O}}_M)^{\dagger} + (\Omega^{\mathcal{O}})^T \Theta_B^{\mathcal{O}} \Omega^{\mathcal{O}},
    \label{eq:GaussianIntegralsGeneralLocalDistributionXi}
\end{equation}
respectively. For practical purposes, we write the latter two in terms of elementary elements using \eqref{eq:GaussianIntegralsLocalCovarianceMatricesRelations}, that is, via the blocks of the covariance matrix, to wit
\begin{equation}
    \begin{split}
        \lambda_A^{\mathcal{O}} &= \theta^{\mathcal{O}} + \Tr_B \left\{ \Theta^{\mathcal{O}}_B \left[ \gamma_{B}^{\mathcal{O}} - (\gamma_{M}^{\mathcal{O}})^T (\gamma_{A}^{\mathcal{O}})^{-1} \gamma_{M}^{\mathcal{O}} \right] \right\}, \\
        \Lambda_A^{\mathcal{O}} &= \Theta_A^{\mathcal{O}} + \Theta_M^{\mathcal{O}} (\gamma^{\mathcal{O}}_M)^T (\gamma^{\mathcal{O}}_A)^{-1} + (\gamma^{\mathcal{O}}_A)^{-1} \gamma^{\mathcal{O}}_M (\Theta_M^{\mathcal{O}} )^{\dagger} + (\gamma^{\mathcal{O}}_A)^{-1} \gamma^{\mathcal{O}}_M \Theta_B^{\mathcal{O}} (\gamma^{\mathcal{O}}_M)^T (\gamma^{\mathcal{O}}_A)^{-1}.
    \end{split}
    \label{eq:GaussianIntegralsGeneralLocalDistributionXiDecomposition}
\end{equation}
Then, putting everything together yields the final result
\begin{equation}
    \mathcal{O}_A [\nu_A] = \frac{1}{Z^{\mathcal{O}}_A} e^{- \frac{1}{2} \nu_A^T (\gamma^{\mathcal{O}}_A)^{-1} \nu_A} \left[ \lambda_A^{\mathcal{O}} + \nu_A^T \Lambda_A^{\mathcal{O}} \nu_A \right],
    \label{eq:GaussianIntegralsGeneralLocalDistribution}
\end{equation}
which is again of the form \eqref{eq:GaussianIntegralsGeneralDistribution}, \textit{i.e.}, a Gaussian times a quadratic form. A well-known identity is obtained for $\lambda_A^{\mathcal{O}} = 1, \Lambda_A^{\mathcal{O}} = 0$ such that $\kappa^{\mathcal{O}} [\nu_A, \nu_B] = 1$: a globally Gaussian distribution is also of Gaussian form in subregion $A$, with the local covariance matrix being the corresponding diagonal block of the global covariance matrix. In this sense, Gaussian distributions are marginalized by simply restricting the global random vector and the global covariance matrix to their local forms.

\subsubsection{Wigner $W$-distribution}
\begin{figure*}[h!]
    \centering
    \includegraphics[width=0.4\textwidth]{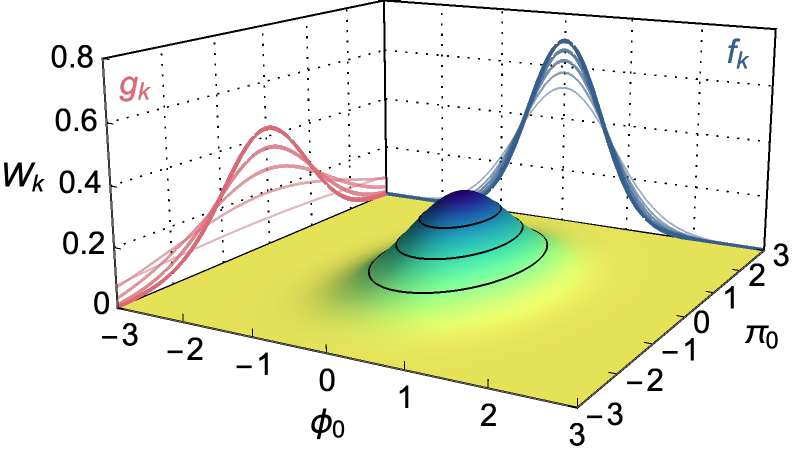}
    \caption{Wigner $W$- (colored surface) and marginal distributions (curves) of the first mode for a particle of momentum $k=50$ and mass $m=1$ on a lattice of spacing $\epsilon = 1$ and length $L=10^2$. All distributions are non-negative and close to Gaussian. Toward the continuum limit, the distributions are squeezed, see thinner lines for $\epsilon=4/5,3/5,2/5,1/5$.}
    \label{fig:DistributionParticlesSM}
\end{figure*}

Since the ground state is Gaussian, we have $\kappa_0^{\mathcal{W}} [\nu_A, \nu_B] = 1$, and thus \eqref{eq:GaussianIntegralsGeneralLocalDistribution} implies
\begin{equation}
    \mathcal{W}_{0,A} [\chi] = \frac{1}{Z_{0,A}^{\mathcal{W}}} e^{- \frac{1}{2} \sum_{j,j' \in \mathfrak{J}_A} \epsilon^2 \chi^T_j (\gamma_{0,A}^{\mathcal{W}})^{-1}_{j j'} \chi_{j'}},
    \label{eq:GroundStateWignerWDistributionPositionSpaceLocal}
\end{equation}
with
\begin{equation}
    (\gamma_{0,A}^{\mathcal{W}})_{j j'} = (\gamma_{0,A}^{f})_{j j'} \oplus (\gamma_{0,A}^{g})_{j j'}, \quad Z_{0,A}^{\mathcal{W}} = \sqrt{\det (2\pi \gamma^{\mathcal{W}}_{0,A} )}.
    \label{eq:GroundStateWignerWCovarianceNormalizationPositionSpaceLocal}
\end{equation}
Therein, the local covariance matrix corresponds to its global counterpart, \textit{i.e.} Eq. \eqref{eq:GroundStateWignerWCovarianceNormalizationPositionSpace}, but with the spatial indices $j,j'$ restricted to the local set $\mathfrak{J}_A$. Similarly, the local normalization constant is obtained by replacing in the global expressions, Eq. \eqref{eq:GroundStateWignerWCovarianceNormalizationPositionSpace}, the global with the local covariance matrix. Let us now consider a quasi-particle. The scalar $\theta_k^{\mathcal{W}}$ and the matrix $\Theta_k^{\mathcal{W}}$ are dictated by \eqref{eq:QuasiParticleWignerWDistributionPositionSpace} and can be formally decomposed in the $\chi = (\phi,\pi)$ basis as
\begin{equation}
    \theta_k^{\mathcal{W}} = -1, \quad (\Theta_k^{\mathcal{W}})_{j j'} = \begin{pmatrix}
        (\Theta_k^{f})_{j j'} & (\Theta_k^{M})_{j j'} \\
        (\Theta_k^{M})^{\dagger}_{j j'} & (\Theta_k^{g})_{j j'}
    \end{pmatrix} = e^{-i \epsilon (j-j') \eta k} \begin{pmatrix}
        \frac{1}{(\tilde{\gamma}_0^f)_{kk}} & - i \frac{2}{L} \\
        i \frac{2}{L} & \frac{1}{(\tilde{\gamma}_0^g)_{kk}}
    \end{pmatrix}.
\end{equation}
To evaluate the local quadratic form, we note again that the off-diagonal blocks describing correlations between the field and the momentum field vanish. This immediately implies a linear decomposition into marginal contributions (the marginal expressions are presented in \autoref{subsubsec:MarginalDistributionsPositionSpaceLocal}) for the scalar part
\begin{equation}
    \lambda_{k,A}^{\mathcal{W}} = -1 + \lambda_{k,A}^f + \lambda_{k,A}^g,
    \label{eq:QuasiParticleWignerxi}
\end{equation}
while the bilinear form also contains off-diagonal terms
\begin{equation}
    (\Lambda^{\mathcal{W}}_{k,A})_{j j'} = \begin{pmatrix}
        (\Lambda^f_{k,A})_{j j'} & (\Lambda^M_{k,A})_{j j'} \\
        (\Lambda^M_{k,A})^{\dagger}_{j j'} & (\Lambda^g_{k,A})_{j j'}
    \end{pmatrix}.
\end{equation}
The off-diagonal block reads
\begin{equation}
    \begin{split}
        (\Lambda^M_{k,A})_{j j'} &= - i \frac{2}{L} \Bigg[ e^{- i \epsilon (j-j') \eta k} + \sum_{j''' \in \mathfrak{J}_A; j'' \in \mathfrak{J}_B} \epsilon^2 \, e^{- i \epsilon (j-j'') \eta k} (\gamma^g_{0,M})^T_{j'' j'''} (\gamma^g_{0,A})^{-1}_{j''' j'} \\
        &\hspace{1.75cm}-\sum_{j'' \in \mathfrak{J}_A; j''' \in \mathfrak{J}_B} \epsilon^2 \, (\gamma^g_{0,A})^{-1}_{j j''} (\gamma^g_{0,M})_{j'' j'''} e^{- i \epsilon (j'''-j') \eta k} \\
        &\hspace{1.75cm}+\sum_{j'', j''''' \in \mathfrak{J}_A; j''', j'''' \in \mathfrak{J}_B} \epsilon^4 \, (\gamma^f_{0,A})^{-1}_{j j''} (\gamma^f_{0,M})_{j'' j'''} e^{i \epsilon (j'''-j'''') \eta k} (\gamma^g_{0,M})^T_{j'''' j'''''} (\gamma^g_{0,A})^{-1}_{j''''' j'} \Bigg],
    \end{split}
    \label{eq:QuasiParticleMixedXi}
\end{equation}
such that, altogether, the local Wigner $W$-distribution of a quasi-particle reads 
\begin{equation}
    \mathcal{W}_{k,A} [\chi] = \mathcal{W}_{0,A} [\chi] \left[ \lambda_{k,A}^{\mathcal{W}} + \sum_{j,j' \in \mathfrak{J}_A} \epsilon^2 \, \chi_j (\Lambda^{\mathcal{W}}_{k,A})_{j j'} \chi_{j'} \right],
    \label{eq:QuasiParticleWignerWDistributionPositionSpaceLocal}
\end{equation}
see \autoref{fig:DistributionParticlesSM} for an example.
For the thermal state, we again have $\kappa_0^{\mathcal{W}} [\nu_A, \nu_B] = 1$ and hence
\begin{equation}
    \mathcal{W}_{T,A} [\chi] = \frac{1}{Z_{T,A}^{\mathcal{W}}} e^{- \frac{1}{2} \sum_{j,j' \in \mathfrak{J}_A} \epsilon^2 \chi^T_j (\gamma_{T,A}^{\mathcal{W}})^{-1}_{j j'} \chi_{j'}},
    \label{eq:ThermalStateWignerWDistributionPositionSpaceLocal}
\end{equation}
where again
\begin{equation}
    (\gamma_{T,A}^{\mathcal{W}})_{j j'} = (\gamma_{T,A}^{f})_{j j'} \oplus (\gamma_{T,A}^{g})_{j j'}, \quad Z_{T,A}^{\mathcal{W}} = \sqrt{\det (2\pi \gamma^{\mathcal{W}}_{T,A})}.
    \label{eq:ThermalStateWignerWCovarianceNormalizationPositionSpaceLocal}
\end{equation}

\subsubsection{Marginal distributions}
\label{subsubsec:MarginalDistributionsPositionSpaceLocal}
The local ground-state marginal distribution over $\phi$ is
\begin{equation}
    f_{0,A} [\phi] = \frac{1}{Z_{0,A}^f} e^{- \frac{1}{2} \sum_{j,j' \in \mathfrak{J}_A} \epsilon^2 \phi_j (\gamma_{0,A}^{f})^{-1}_{j j'} \phi_{j'}},
    \label{eq:GroundStateMarginalDistributionPositionSpaceLocal}
\end{equation}
with the local covariance matrix following from \eqref{eq:GroundStateMarginalCovarianceNormalizationPositionSpace} by restricting the indices to $j,j' \in \mathfrak{J}_A$ and $Z_{0,A}^f = \sqrt{\det (2\pi \gamma_{0,A}^f)}$. For a quasi-particle, we consider Eq. \eqref{eq:QuasiParticleMarginalDistributionPositionSpace} to read-off the quadratic form
\begin{equation}
    (\Theta^{f}_k)_{j j'} = \frac{e^{- i \epsilon (j-j') \eta k}}{(\tilde{\gamma}_0^f)_{k k}}.
\end{equation}
This together with \eqref{eq:GaussianIntegralsGeneralLocalDistributionXi} implies for the two building blocks of the local quadratic form in \eqref{eq:GaussianIntegralsGeneralLocalDistribution}
\begin{equation}
    \begin{split}
        \lambda^f_{k,A} &= \frac{1}{(\tilde{\gamma}_0^f)_{k k}} \sum_{j,j' \in \mathfrak{J}_B} \epsilon^2 \, e^{- i \epsilon (j-j') \eta k} \Bigg[ (\gamma_{0,B}^f)_{j'j} - \sum_{j'',j''' \in \mathfrak{J}_A}\epsilon^2 (\gamma_{0,M}^f)^T_{j' j''} (\gamma_{0,A}^f)^{-1}_{j''j'''} (\gamma_{0,M}^f)_{j'''j} \Bigg], \\
        (\Lambda^f_{k,A})_{j j'} &= \frac{1}{(\tilde{\gamma}_0^f)_{k k}} \Bigg[ e^{- i \epsilon (j-j') \eta k} + \sum_{j''' \in \mathfrak{J}_A; j'' \in \mathfrak{J}_B} \epsilon^2 \, e^{- i \epsilon (j-j'') \eta k} (\gamma^f_{0,M})^T_{j'' j'''} (\gamma^f_{0,A})^{-1}_{j''' j'} \\
        &\hspace{1.75cm}+\sum_{j'' \in \mathfrak{J}_A; j''' \in \mathfrak{J}_B} \epsilon^2 \, (\gamma^f_{0,A})^{-1}_{j j''} (\gamma^f_{0,M})_{j'' j'''} e^{- i \epsilon (j'''-j') \eta k} \\
        &\hspace{1.75cm}+\sum_{j'', j''''' \in \mathfrak{J}_A; j''', j'''' \in \mathfrak{J}_B} \epsilon^4 \, (\gamma^f_{0,A})^{-1}_{j j''} (\gamma^f_{0,M})_{j'' j'''} e^{i \epsilon (j'''-j'''') \eta k} (\gamma^f_{0,M})^T_{j'''' j'''''} (\gamma^f_{0,A})^{-1}_{j''''' j'} \Bigg],
    \end{split}
    \label{eq:QuasiParticleMarginalXi}
\end{equation}
where $j,j' \in \mathfrak{J}_A$, such that (see \autoref{fig:DistributionParticlesSM})
\begin{equation}
    f_{k,A} [\phi] = f_{0,A} [\phi] \left[ \lambda_{k,A}^f + \sum_{j,j' \in \mathfrak{J}_A} \epsilon^2 \, \phi_j (\Lambda^{f}_{k,A})_{j j'} \phi_{j'} \right].
    \label{eq:QuasiParticleMarginalDistributionPositionSpaceLocal}
\end{equation}
Finally, the local distribution for the thermal state reads
\begin{equation}
    f_{T,A} [\phi] = \frac{1}{Z_{T,A}^f} e^{- \frac{1}{2} \sum_{j,j' \in \mathfrak{J}_A} \epsilon^2 \phi_j (\gamma_{T,A}^{f})^{-1}_{j j'} \phi_{j'}},
    \label{eq:ThermalStateMarginalDistributionPositionSpaceLocal}
\end{equation}
with the covariance matrix given in Eq. \eqref{eq:ThermalStateMarginalCovariancePositionSpace} restricted to $j,j' \in \mathfrak{J}_A$ and $Z^f_{T,A} = \sqrt{\det (2 \pi \gamma_{T,A}^f)}$. The corresponding expressions for $g_A [\pi]$ follow analogously.

\subsubsection{Husimi $Q$-distribution}
The Husimi $Q$-distribution of the ground state is
\begin{equation}
    \mathcal{Q}_{0,A} [\chi] = \frac{1}{Z_{0,A}^{\mathcal{Q}}} e^{- \frac{1}{2} \sum_{j,j' \in \mathfrak{J}_A} \epsilon^2 \chi^T_j (\gamma_{0,A}^{\mathcal{Q}})^{-1}_{j j'} \chi_{j'}},
    \label{eq:GroundStateHusimiQDistributionPositionSpaceLocal}
\end{equation}
with the local covariance matrix following from \eqref{eq:GroundStateHusimiQCovarianceNormalizationPositionSpace} and normalization $Z_{0,A}^{\mathcal{Q}} = \sqrt{\det ( 2 \pi\gamma_{0,A}^{\mathcal{Q}})}$. For a quasi-particle, we compare \eqref{eq:QuasiParticleHusimiQDistributionPositionSpace} with the corresponding expression for the Wigner $W$-distribution \eqref{eq:QuasiParticleWignerWDistributionPositionSpace}, which shows that all submatrices of the quadratic form of the former, \textit{i.e.}, 
\begin{equation}
    (\Theta_k^{\mathcal{Q}})_{j j'} = \begin{pmatrix}
        (\Theta_k^{\mathcal{Q},f})_{j j'} & (\Theta_k^{\mathcal{Q},M})_{j j'} \\
        (\Theta_k^{\mathcal{Q},M})^{\dagger}_{j j'} & (\Theta_k^{\mathcal{Q},g})_{j j'}
    \end{pmatrix},
\end{equation}
are proportional to the latter in the sense that
\begin{equation}
    \Theta_k^{\mathcal{Q},f} = \left[\frac{(\tilde{\gamma}^f_0)_{kk}}{(\tilde{\gamma}^f_0)_{kk} + \bar{\tilde{\gamma}}^f_{k k}} \right]^2 \Theta_k^{f}, \quad \Theta_k^{\mathcal{Q},M} = \frac{L^2}{4} \frac{\Theta_k^{M}}{\left[(\tilde{\gamma}^f_0)_{kk} +\bar{\tilde{\gamma}}^{f}_{k k} \right]\left[(\tilde{\gamma}^g_0)_{kk} + \bar{\tilde{\gamma}}^{g}_{k k} \right]}, \quad \Theta_k^{\mathcal{Q},g} = \left[\frac{(\tilde{\gamma}^g_0)_{kk}}{(\tilde{\gamma}^g_0)_{kk} + \bar{\tilde{\gamma}}^g_{k k}} \right]^2 \Theta_k^{g}.
\end{equation}
This implies the decompositions
\begin{equation}
    \lambda_{k,A}^{\mathcal{Q}} = \lambda_{k,A}^{\mathcal{Q},f} + \lambda_{k,A}^{\mathcal{Q},g}, \quad 
    (\Lambda^{\mathcal{Q}}_{k,A})_{j j'} = \begin{pmatrix}
        (\Lambda^{\mathcal{Q},f}_{k,A})_{j j'} & (\Lambda^{\mathcal{Q},M}_{k,A})_{j j'} \\
        (\Lambda^{\mathcal{Q},M}_{k,A})^{\dagger}_{j j'} & (\Lambda^{\mathcal{Q},g}_{k,A})_{j j'},
    \end{pmatrix},
    \label{eq:QuasiParticleHusimixiXi}
\end{equation}
with the building blocks of the final quadratic form being form-equivalent to \eqref{eq:QuasiParticleMarginalXi} and \eqref{eq:QuasiParticleMixedXi}, but with the marginal covariances replaced by the corresponding blocks of the Husimi $Q$'s covariance matrix, and adjusted prefactors, to wit
\begin{equation}
    \begin{split}
        \lambda^{\mathcal{Q},f}_{k,A} &= \frac{(\tilde{\gamma}^f_0)_{kk}}{\left[(\tilde{\gamma}^f_0)_{kk} + \bar{\tilde{\gamma}}^{f}_{k k} \right]^2} \sum_{j,j' \in \mathfrak{J}_B} \epsilon^2 \, e^{- i \epsilon (j-j') \eta k} \Bigg[ (\gamma_{0,B}^{\mathcal{Q},f})_{j'j} - \sum_{j'',j''' \in \mathfrak{J}_A}\epsilon^2 (\gamma_{0,M}^{\mathcal{Q},f})^T_{j' j''} (\gamma_{0,A}^{\mathcal{Q},f})^{-1}_{j''j'''} (\gamma_{0,M}^{\mathcal{Q},f})_{j'''j} \Bigg], \\
        (\Lambda^{\mathcal{Q}, f}_{k,A})_{j j'} &= \frac{(\tilde{\gamma}^f_0)_{kk}}{\left[(\tilde{\gamma}^f_0)_{kk} + \bar{\tilde{\gamma}}^{f}_{k k} \right]^2} \Bigg[ e^{- i \epsilon (j-j') \eta k} + 2 \sum_{j''' \in \mathfrak{J}_A; j'' \in \mathfrak{J}_B} \epsilon^2 \, \cos \left[ \epsilon (j-j'') \eta k \right] (\gamma^{\mathcal{Q},f}_{0,M})^T_{j'' j'''} (\gamma^{\mathcal{Q},f}_{0,A})^{-1}_{j''' j'} \\
        &\hspace{1.75cm}+\sum_{j'', j''''' \in \mathfrak{J}_A; j''', j'''' \in \mathfrak{J}_B} \epsilon^4 \, (\gamma^{\mathcal{Q},f}_{0,A})^{-1}_{j j''} (\gamma^{\mathcal{Q},f}_{0,M})_{j'' j'''} e^{i \epsilon (j'''-j'''') \eta k} (\gamma^{\mathcal{Q},f}_{0,M})^T_{j'''' j'''''} (\gamma^{\mathcal{Q},f}_{0,A})^{-1}_{j''''' j'} \Bigg],
    \end{split}
    \label{eq:QuasiParticleHusimixiXi2}
\end{equation}
(analogously for the quantities over the momentum field $\pi$ and the off-diagonal block $\Lambda_{0,A}^{\mathcal{Q},M}$). This results in 
\begin{equation}
    \mathcal{Q}_{k,A} [\chi] = \mathcal{Q}_{0,A} [\chi] \left[ \lambda_{k,A}^{\mathcal{Q}} + \sum_{j,j' \in \mathfrak{J}_A} \epsilon^2 \, \chi_j (\Lambda^{\mathcal{Q}}_{k,A})_{j j'} \chi_{j'} \right],
    \label{eq:QuasiParticleHusimiQDistributionPositionSpaceLocal}
\end{equation}
while for the thermal state, we have
\begin{equation}
    \mathcal{Q}_{T,A} [\chi] = \frac{1}{Z_{T,A}^{\mathcal{Q}}} e^{- \frac{1}{2} \sum_{j,j' \in \mathfrak{J}_A} \epsilon^2 \chi^T_j (\gamma_{T,A}^{\mathcal{Q}})^{-1}_{j j'} \chi_{j'}},
    \label{eq:ThermalStateHusimiQDistributionPositionSpaceLocal}
\end{equation}
with Eq. \eqref{eq:ThermalStateHusimiQCovarianceNormalizationPositionSpace} understood and $Z_{T,A}^{\mathcal{Q}} = \sqrt{\det ( 2 \pi \gamma_{T,A}^{\mathcal{Q}})}$.

\section{Subtracted entropies}
\label{sec:SubtractedEntropies}
Here, we derive the general formula for the subtracted Rényi entropy of the local distributions $\mathcal{O}_A$ of our interest. To this end, we apply the central identity of quantum field theory once again, resulting in a closed expression that can be evaluated diagrammatically. Thereupon, we provide more details on why the specific terms discussed in the main text are subtracted to end up with well-behaved entropic measures.

\subsection{General formulae}

\subsubsection{Intermezzo on Gaussian integrals II}
\label{subsubsec:GaussianIntegralsII}
We start with stating the central identity of quantum field theory for the subregion $A$. Following the conventions and results of \autoref{subsubsec:GaussianIntegralsI}, we consider a distribution over the $v$-dimensional random vector $\nu_A$ of the form
\begin{equation}
    \mathcal{O}_A [\nu_A] = G^{\mathcal{O}}_A [\nu_A] \, \kappa^{\mathcal{O}}_A [\nu_A].
    \label{eq:GaussianIntegralsGeneralDistributionExponent}
\end{equation}
Therein, $G^\mathcal{O}_A [\nu_A]$ is again a Gaussian distribution
\begin{equation}
    G^\mathcal{O}_A [\nu_A] = \frac{1}{Z^{\mathcal{O}}_A} \, e^{- \frac{1}{2} \nu_A^T (\gamma^{\mathcal{O}}_A)^{-1} \nu_A},
\end{equation}
with zero mean, covariance matrix $\gamma^{\mathcal{O}}_A$ and normalization $Z^{\mathcal{O}}_A = \sqrt{\det (2\pi \gamma^{\mathcal{O}}_A)}$ and $\kappa^{\mathcal{O}}_A [\nu]$ is still a quadratic form, \textit{i.e.},
\begin{equation}
    \kappa^{\mathcal{O}}_A [\nu_A] = \lambda_A^{\mathcal{O}} + \nu_A^T \Lambda_A^{\mathcal{O}} \nu_A.
\end{equation}
We ultimately wish to compute the integral over the distribution $\mathcal{O}_A [\nu_A]$ raised to an arbitrary power $r$, namely
\begin{equation}
    \int \mathcal{D} \nu_A \, \mathcal{O}^r_A [\nu_A] = \frac{1}{(Z^{\mathcal{O}}_A)^r} \int \mathcal{D} \nu_A \, e^{- \frac{r}{2} \nu_A^T (\gamma^{\mathcal{O}}_A)^{-1} \nu_A} \left(\kappa_A^{\mathcal{O}} [\nu_A] \right)^r.
\end{equation}
Analogous to the procedure in \autoref{subsubsec:GaussianIntegralsI}, we introduce a source $\zeta$ to rewrite the latter as 
\begin{equation}
    \int \mathcal{D} \nu_A \, \mathcal{O}_A^r [\nu_A] = \left(\kappa_A^{\mathcal{O}} [\partial_{\zeta}]\right)^r \frac{1}{(Z_A^{\mathcal{O}})^r} \int \mathcal{D} \nu_A \, e^{- \frac{r}{2} \nu_A^T (\gamma^{\mathcal{O}}_A)^{-1} \nu_A + \zeta^T \nu_A} \Big \vert_{\zeta = 0}.
\end{equation}
The remaining integral is a standard Gaussian integral with respect to the rescaled covariance matrix $\gamma^{\mathcal{O}}_A/r$, which can be evaluated straightforwardly
\begin{equation}
    \frac{1}{(Z_A^{\mathcal{O}})^r} \int \mathcal{D} \nu_A \, e^{- \frac{r}{2} \nu_A^T (\gamma^{\mathcal{O}}_A)^{-1} \nu_A + \zeta^T \nu_A} = \frac{\sqrt{\det \left[2 \pi (\gamma^{\mathcal{O}}_A / r) \right]}}{(Z^{\mathcal{O}}_A)^r} \, e^{\frac{1}{2 r} \zeta^T \gamma^{\mathcal{O}}_A \zeta} = \sqrt{\frac{\text{det}^{(1-r)} (2 \pi \gamma_A^{\mathcal{O}})}{r^{\dim (\gamma_A^{\mathcal{O}})}}} \, e^{\frac{1}{2 r} \zeta^T \gamma_A^{\mathcal{O}} \zeta},
\end{equation}
and therefore
\begin{equation}
     \int \mathcal{D} \nu_A \, \mathcal{O}_A^r [\nu_A] = \sqrt{\frac{\text{det}^{(1-r)} (2 \pi \gamma_A^{\mathcal{O}})}{r^{\dim (\gamma_A^{\mathcal{O}})}}} \, U_A^{\mathcal{O}} (r),    \label{eq:CentralIdentityQFT}
\end{equation}
where we introduced the real number
\begin{equation}
    U_A^{\mathcal{O}} (r) = \left(\kappa_A^{\mathcal{O}} [\partial_{\zeta}]\right)^r e^{\frac{1}{2 r} \zeta^T \gamma^{\mathcal{O}}_A \zeta} \Big \vert_{\zeta = 0}.
\end{equation}
Next, we have to calculate some derivatives, for which we shall assume $r \in \mathbb{N}$ to be a natural number (this is reminiscent of the replica method, a well-known trick in quantum field theory to compute the entanglement entropy). Since $\kappa^{\mathcal{O}}_A [\nu]$ is a sum of two terms we apply the binomial theorem
\begin{equation}
    \left(\kappa_A^{\mathcal{O}} [\nu_A] \right)^r = \left[ \lambda_A^{\mathcal{O}} + \nu_A^T \Lambda_A^{\mathcal{O}} \nu_A \right]^r = \sum_{s=0}^r \begin{pmatrix}
            r \\ s
        \end{pmatrix} 
        (\lambda_A^{\mathcal{O}})^{r-s} \left( \nu_A^T \Lambda_A^{\mathcal{O}} \nu_A \right)^s,
\end{equation}
such that
\begin{equation}
    U^{\mathcal{O}}_A (r) = \sum_{s=0}^r \begin{pmatrix}
            r \\ s
        \end{pmatrix} 
        (\lambda_A^{\mathcal{O}})^{r-s} \left( \partial_{\zeta}^T \Lambda_A^{\mathcal{O}} \partial_{\zeta} \right)^s e^{\frac{1}{2 r} \zeta^T \gamma^{\mathcal{O}}_A \zeta} \Big \vert_{\zeta = 0}.
\end{equation}
We remark that a generalization to real $r$ at this point requires the usage of the generalized binomial theorem, in which case the binomial coefficient is replaced by a fraction of Gamma functions and the sum runs up to infinity for non-integer $r$. When expanding the exponential, the only term surviving at $\zeta = 0$ is the term of order $2s$ as all other terms are multiplied by some power of $\zeta$ and thus
\begin{equation}
     \left( \partial_{\zeta}^T \Lambda_A^{\mathcal{O}} \partial_{\zeta} \right)^s e^{\frac{1}{2 r} \zeta^T \gamma_A^{\mathcal{O}} \zeta} \Big \vert_{\zeta = 0} = \frac{1}{r^s} \frac{1}{2^s s!} \left( \partial_{\zeta}^T \Lambda_A^{\mathcal{O}} \partial_{\zeta} \right)^s \left( \zeta^T \gamma_A^{\mathcal{O}} \zeta \right)^s \equiv \frac{u_A^{\mathcal{O}} (s)}{r^s}.
\end{equation}

The remaining problem is to determine $u_A^{\mathcal{O}} (s)$, which is deeply rooted in combinatorics and is of similar form as the problems appearing in the proofs of the well-known theorems by Isserlis and Wick. To keep track of the multiplicity of all terms properly, we write out all inner products in terms of their components, i.e. 
\begin{equation}
    u_A^{\mathcal{O}} (s) = \frac{1}{2^s s!} \sum_{i_n, i'_n} \left[ (\Lambda_A^{\mathcal{O}})_{i_1 i_2} \dots (\Lambda_A^{\mathcal{O}})_{i_{2s-1} i_{2s}} \right] \left[ (\gamma_A^{\mathcal{O}})_{i'_1 i'_2} \dots (\gamma_A^{\mathcal{O}})_{i'_{2s - 1} i'_{2s}} \right] \left(\partial_{\zeta_{i_1}} \dots \partial_{\zeta_{i_{2s}}} \right) \left(\zeta_{i'_1} \dots \zeta_{i'_{2s} }\right).
\end{equation}
Every derivative with respect to a component of $\zeta$ can act on any other component of $\zeta$ appearing in the quadratic form with respect to $\gamma^{\mathcal{O}}_A$, such that there are $(2s)!$ permutations for how to evaluate the derivatives. Hence,
\begin{equation}
    \frac{1}{2^s s!} \sum_{i'_n} \left[ (\gamma_A^{\mathcal{O}})_{i'_1 i'_2} \dots (\gamma_A^{\mathcal{O}})_{i'_{2s - 1} i'_{2s}} \right] \left(\partial_{\zeta_{i_1}} \dots \partial_{\zeta_{i_{2s}}} \right) \left(\zeta_{i'_1} \dots \zeta_{i'_{2s} }\right) = \frac{1}{2^s s!} \sum_{\sigma \in S_{2s}} \left[ (\gamma_A^{\mathcal{O}})_{i_{\sigma(1)} i_{\sigma(2)}} \dots (\gamma_A^{\mathcal{O}})_{i_{\sigma(2s-1)} i_{\sigma(2s)}} \right], 
\end{equation}
wherein $S_{2s}$ denotes the symmetric group of order $2s$. It is important to note here that the above sum is degenerate in two ways: First, $\gamma^{\mathcal{O}}_A$ is a covariance matrix and as such symmetric, leading to a degeneracy of order $2^s$. Second, the order of the terms does not matter, giving another degeneracy of order $s!$. As a result, there appear only
\begin{equation}
    (2s-1)!! \equiv \frac{(2s)!}{2^s s!}
\end{equation}
independent terms in the above sum, where $!!$ defines the double factorial over an odd number. This leads us to the essence of the two aforementioned theorems, which is the simplification
\begin{equation}
    \begin{split}
        u_A^{\mathcal{O}} (s) = \sum_{i_n} \sum_{\sigma \in P^2_{2s}} \left[ (\Lambda_A^{\mathcal{O}})_{i_1 i_2} \dots (\Lambda_A^{\mathcal{O}})_{i_{2s-1} i_{2s}} \right] \left[ (\gamma_A^{\mathcal{O}})_{i_{\sigma(1)} i_{\sigma(2)}} \dots (\gamma_A^{\mathcal{O}})_{i_{\sigma(2s-1)} i_{\sigma(2s)}} \right],
    \end{split}
    \label{eq:GaussianIntegralsIIu}
\end{equation}    
where $P^2_{2s}$ is the set of all distinct partitions of the set $\{1, ..., 2s \}$ into pairs, \textit{i.e.}, subsets of order $2$, which indeed contains only $(2s-1)!!$ elements. On the right-hand side, we are left with a sum over powers of the trace of the matrix products $(\Lambda_A^{\mathcal{O}} \gamma_A^{\mathcal{O}})^t$ and $(\gamma_A^{\mathcal{O}} \Lambda_A^{\mathcal{O}})^t$, where $t \le s$ is yet another power. We already here note that the cyclicity of the trace as well as its invariance under transposition further reduce the number of independent terms in Eq. \eqref{eq:GaussianIntegralsIIu}. In \autoref{subsubsec:DiagramsNotation}, we present a simple graphical method to calculate $u_A^{\mathcal{O}} (s)$ and in particular the degeneracies of all terms involved, which we will carry out for the first four values of $s$ in \autoref{subsubsec:LowestOrderTerms}. For now, we just put everything together, which results in the closed formula
\begin{equation}
    \int \mathcal{D} \nu_A \, \mathcal{O}^r [\nu_A] = \sqrt{\frac{\text{det}^{(1-r)} (2 \pi \gamma_A^{\mathcal{O}})}{r^{\dim (\gamma_A^{\mathcal{O}})}}} \, U_A^{\mathcal{O}} (r) = \sqrt{\frac{\text{det}^{(1-r)} (2 \pi \gamma_A^{\mathcal{O}})}{r^{\dim (\gamma_A^{\mathcal{O}})}}} \, \sum_{s=0}^r \begin{pmatrix}
            r \\ s
        \end{pmatrix} 
        \frac{(\lambda_A^{\mathcal{O}})^{r-s}}{r^s} \, u_A^{\mathcal{O}} (s).    \label{eq:GaussianIntegralsIIGeneralPower}
\end{equation}

\subsubsection{$u_A^{\mathcal{O}} (s)$ from diagrams}
\label{subsubsec:DiagramsNotation}
The task of calculating $u_A^{\mathcal{O}} (s)$ boils down to an exercise in combinatorics, which can become tedious when $s$ grows large. Instead of pursuing a pen and paper approach, we suggest a simple diagrammatical method. 

\begin{figure}[h!]
    \centering
    \includegraphics[width=0.95\columnwidth]{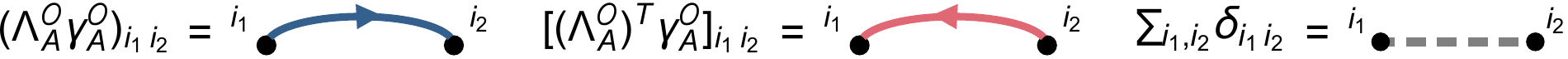}
    \caption{Diagrammatic notation for the three types of terms appearing in Eq. \eqref{eq:GaussianIntegralsIIu}.}
    \label{fig:DiagramNotation}
\end{figure}

Upon inspecting Eq. \eqref{eq:GaussianIntegralsIIu}, we can identify three elementary building blocks: components of the matrix products $\Lambda_A^{\mathcal{O}} \gamma_A^{\mathcal{O}}$ and $(\Lambda_A^{\mathcal{O}})^T \gamma_A^{\mathcal{O}} = (\gamma_A^{\mathcal{O}} \Lambda_A^{\mathcal{O}})^T$ as well as sums over these components. We depict both matrix products by two black points, representing the two indices of the matrix, connected by a curved arrow, which is blue (red) and points from the first (second) to the second (first) index when the matrix product $\Lambda_A^{\mathcal{O}} \gamma_A^{\mathcal{O}}$ ($(\Lambda_A^{\mathcal{O}})^T \gamma_A^{\mathcal{O}}$) is considered, see the two left diagrams in \autoref{fig:DiagramNotation}. Further, we sketch the sum over two identified indices by connecting two black dots by a dashed, straight line (right diagram in \autoref{fig:DiagramNotation}). 

In this notation, $u_A^{\mathcal{O}} (s)$ is nothing but the sum over all independent closed and directed graphs connecting all matrix-product endpoints in the form of closed loops such that all vertices have two edges. When a diagram can be represented by two or more closed graphs, the corresponding term is multiplied by the diagram's degeneracy.

Based on our earlier considerations, there are $(2s-1)!!$ diagrams for a given order $s$, but we will see below that not all of them are independent since the trace is invariant under symmetric permutations as well as transposition. As also the direction of the graph is irrelevant, we always consider the indices $i_1$ and $i_2$ to be connected by a blue arrow (thereby representing the matrix product $\Lambda_A^{\mathcal{O}} \gamma_A^{\mathcal{O}}$) when drawing a graph, which we then connect to open indices of the remaining matrix products.

\subsubsection{Lowest-order terms}
\label{subsubsec:LowestOrderTerms}
\begin{figure}[h!]
    \centering
    \includegraphics[width=0.47\columnwidth]{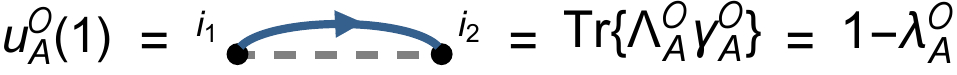}
    \caption{The only diagram for $s=1$.}
    \label{fig:Diagrams1}
\end{figure}

We are now ready to give explicit expressions for $u_A^{\mathcal{O}} (s)$ and $U_A^{\mathcal{O}} (r)$. We start with $r=1$, in which case the normalization of $\mathcal{O}_A [\nu_A]$ introduces the constraint
\begin{equation}
     1 = \int \mathcal{D} \nu_A \, \mathcal{O}_A [\nu_A] = U_A^{\mathcal{O}} (1) = \lambda_A^{\mathcal{O}} + u_A^{\mathcal{O}} (1) =  \lambda_A^{\mathcal{O}} + \Tr \{ \Lambda_A^{\mathcal{O}} \gamma_A^{\mathcal{O}} \},
     \label{eq:GaussianIntegralsIIU1u1}
\end{equation}
which will be of great use in simplifying the higher-order expressions. The only diagram for $s=1$ is sketched in \autoref{fig:Diagrams1} and corresponds to a trace over the matrix product $\Lambda_A^{\mathcal{O}} \gamma_A^{\mathcal{O}}$ (blue dashed line) since the sum runs over both matrix indices. Note also that $(2s-1)!! = 1$ for $s=1$, as it should be.

\begin{figure}[h!]
    \centering
    \includegraphics[width=0.63\columnwidth]{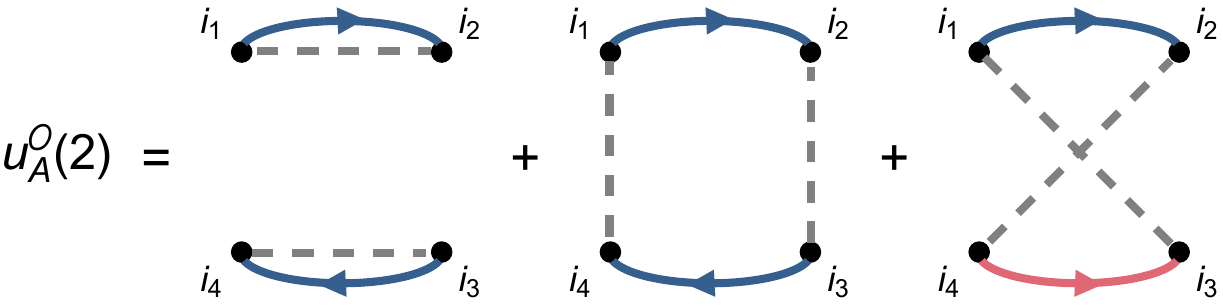}
    \caption{The three diagrams for $s=2$.}
    \label{fig:Diagrams2}
\end{figure}

The first non-trivial case is $r=2$, for which
\begin{equation}
    U_A^{\mathcal{O}} (2) = (\lambda_A^{\mathcal{O}})^2 + \lambda_A^{\mathcal{O}} u_A^{\mathcal{O}} (1) + \frac{u_A^{\mathcal{O}} (2)}{4}.
\end{equation}
Therein, $u_A^{\mathcal{O}} (1)$ can be expressed in terms of $\lambda_A^{\mathcal{O}}$ only using \eqref{eq:GaussianIntegralsIIU1u1}, whereas $u_A^{\mathcal{O}} (2)$ evaluates to the $(2 \cdot 2  -1)!! = 3$ diagrams shown in \autoref{fig:Diagrams2}. 
Thus,
\begin{equation}
    u_A^{\mathcal{O}} (2) = (1 - \lambda_A^{\mathcal{O}})^2 + \Tr \{ (\Lambda_A^{\mathcal{O}} \gamma_A^{\mathcal{O}})^2 \} + \Tr \{ \Lambda_A^{\mathcal{O}} \gamma_A^{\mathcal{O}} (\Lambda_A^{\mathcal{O}})^T \gamma_A^{\mathcal{O}} \},
    \label{eq:GaussianIntegralsIIu2}
\end{equation}
and after plugging the latter equation into the former, we obtain the simple formula
\begin{equation}
    U_A^{\mathcal{O}} (2) = \left( \frac{1 + \lambda_A^{\mathcal{O}}}{2} \right)^2 + \frac{1}{4} \Tr \{ \Lambda_A^{\mathcal{O}} \gamma_A^{\mathcal{O}} \left[ (\Lambda_A^{\mathcal{O}})^T + \Lambda_A^{\mathcal{O}} \right] \gamma_A^{\mathcal{O}} \}.
    \label{eq:GaussianIntegralsIIU2}
\end{equation}

\begin{figure}[h!]
    \centering
    \includegraphics[width=0.73\columnwidth]{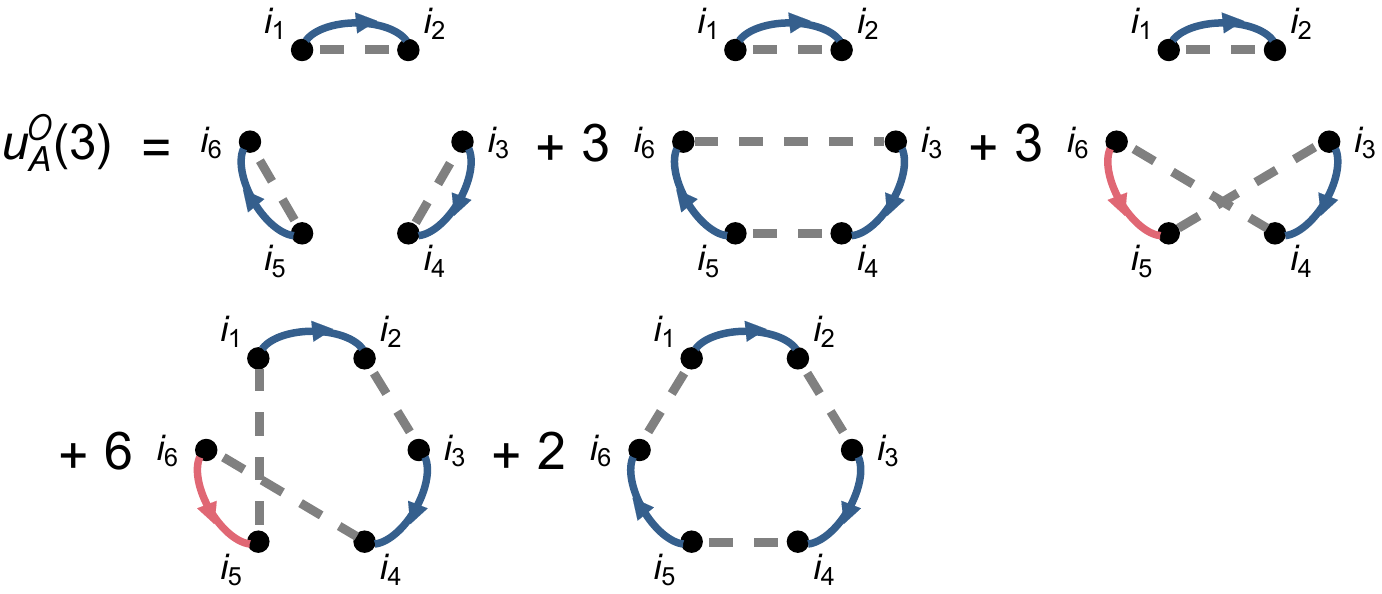}
    \caption{The five diagrams for $s=3$ together with their degeneracies.}
    \label{fig:Diagrams3}
\end{figure}

Further, for $r=3$, we have
\begin{equation}
    U_A^{\mathcal{O}} (3) = (\lambda_A^{\mathcal{O}})^3 + (\lambda_A^{\mathcal{O}})^2 u_A^{\mathcal{O}} (1) + \frac{\lambda_A^{\mathcal{O}} u_A^{\mathcal{O}}(2)}{3} + \frac{u_A^{\mathcal{O}}(3)}{27}.
\end{equation}
Let us discuss the corresponding diagrams for $s=3$ (see \autoref{fig:Diagrams3}) in detail. In contrast to $s=1$ and $s=2$, not all $(2 \cdot 3 -1)!! = 15$ diagrams are independent. While the first diagram in \autoref{fig:Diagrams3} is unique (this holds true for arbitrary $s$), the second and third diagrams carry a degeneracy of three since both larger loops can also run over $i_1 \to i_2 \to i_4 \to i_3$ or $i_1 \to i_2 \to i_6 \to i_5$. As a consequence of the invariance of the trace under transposition, the fourth diagram is even more degenerate, since additionally to connecting $i_1$ to either $i_3, i_4, i_5$ or $i_6$ and having one red arrow in the loop, one may also connect $i_1$ to $i_3$ or $i_5$ with two red arrows. For the last term, we only have two possibilities: either connecting all indices in an outer loop (which is shown in \autoref{fig:Diagrams3}), or, connecting $i_2$ to $i_5$ and then $i_6$ to $i_4$, which gives a star-like graph. Therefore, we get in total
\begin{equation}
    \begin{split}
        u_A^{\mathcal{O}} (3) &= (1 - \lambda_A^{\mathcal{O}})^3 + 3 (1 - \lambda_A^{\mathcal{O}}) \Tr \{ \Lambda_A^{\mathcal{O}} \gamma_A^{\mathcal{O}} (\Lambda_A^{\mathcal{O}})^T \gamma_A^{\mathcal{O}} \} + 3 (1 - \lambda_A^{\mathcal{O}}) \Tr \{ (\Lambda_A^{\mathcal{O}} \gamma_A^{\mathcal{O}})^2 \} \\
        &\hspace{0.4cm}+ 6 \Tr \{ (\Lambda_A^{\mathcal{O}} \gamma_A^{\mathcal{O}})^2 (\Lambda_A^{\mathcal{O}})^T \gamma_A^{\mathcal{O}} \} + 2 \Tr \{ (\Lambda_A^{\mathcal{O}} \gamma_A^{\mathcal{O}})^3 \},
    \end{split}
    \label{eq:GaussianIntegralsIIu3}
\end{equation}
and, after a few algebraic manipulations, we end up with
\begin{equation}
    U_A^{\mathcal{O}} (3) = \left( \frac{1 + 2 \lambda_A^{\mathcal{O}}}{3} \right)^3 + \frac{1 + 2 \lambda_A^{\mathcal{O}}}{9} \Tr \{ \Lambda_A^{\mathcal{O}} \gamma_A^{\mathcal{O}} \left[ (\Lambda_A^{\mathcal{O}})^T + \Lambda_A^{\mathcal{O}} \right] \gamma_A^{\mathcal{O}} \} + \frac{2}{27} \Tr \left\{ (\Lambda_A^{\mathcal{O}} \gamma_A^{\mathcal{O}})^2 \left[3 (\Lambda_A^{\mathcal{O}})^T + \Lambda_A^{\mathcal{O}} \right] \gamma_A^{\mathcal{O}} \right \}.
    \label{eq:GaussianIntegralsIIU3}
\end{equation}

\begin{figure}[h!]
    \centering
    \includegraphics[width=0.78\columnwidth]{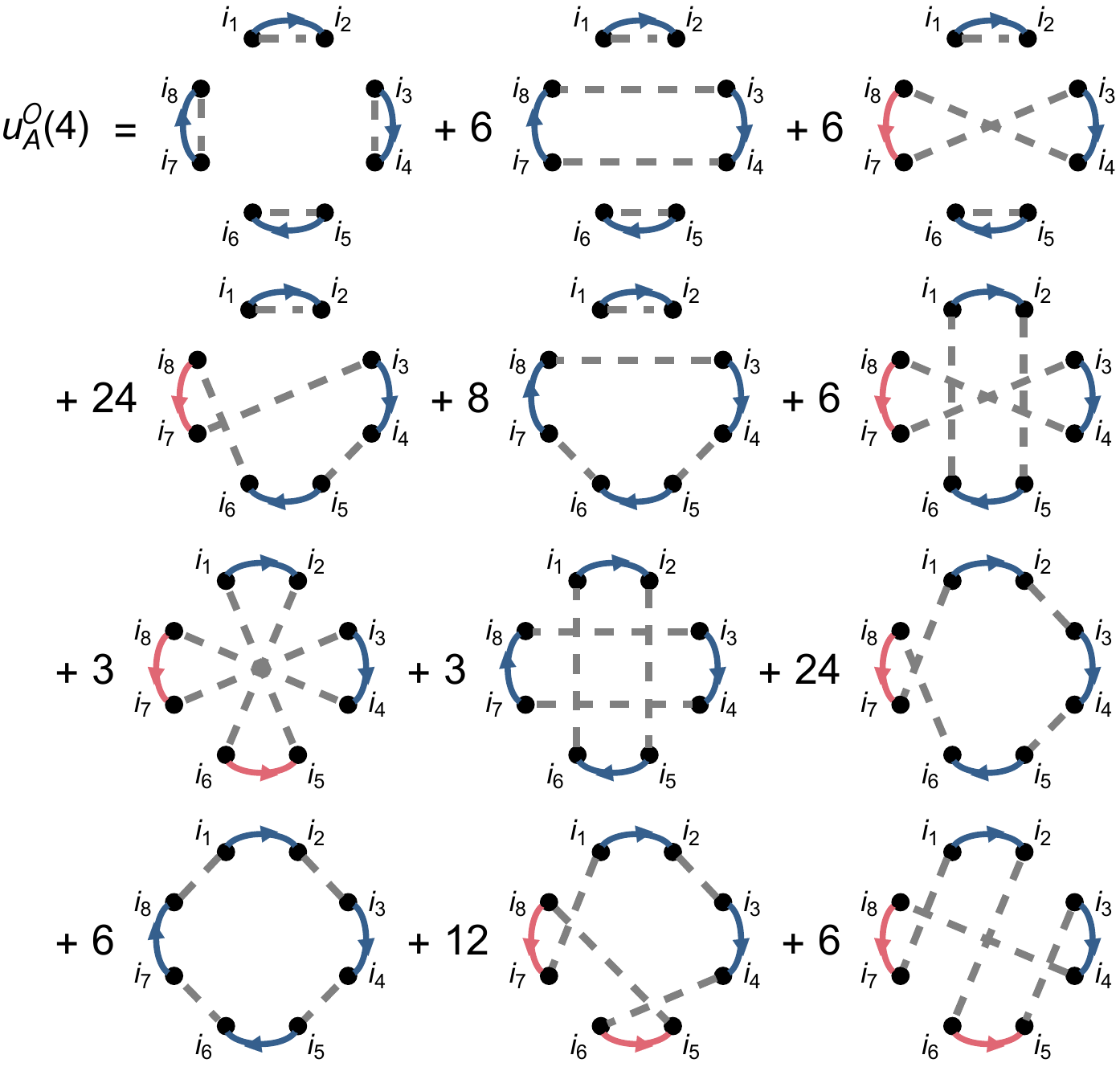}
    \caption{The twelve independent diagrams for $s=4$.}
    \label{fig:Diagrams4}
\end{figure}

At last, we explicitly consider $r=4$, for which
\begin{equation}
    \begin{split}
        U_A^{\mathcal{O}} (4) &= (\lambda_A^{\mathcal{O}})^4 + (\lambda_A^{\mathcal{O}})^3 u_A^{\mathcal{O}} (1) + \frac{3}{8} (\lambda_A^{\mathcal{O}})^2 u_A^{\mathcal{O}} (2) + \frac{\lambda_A^{\mathcal{O}} u_A^{\mathcal{O}} (3)}{16} + \frac{u_A^{\mathcal{O}} (4)}{256}.
    \end{split} 
\end{equation}
For $s=4$, we have $(2 \cdot 4 - 1)!! = 105$ diagrams in total, of which 12 are independent (see \autoref{fig:Diagrams4}). We find
\begin{equation}
    \begin{split}
        u_A^{\mathcal{O}} (4) &= (1 - \lambda_A^{\mathcal{O}})^4 + 6 (1 - \lambda_A^{\mathcal{O}})^2 \Tr \{ (\Lambda_A^{\mathcal{O}} \gamma_A^{\mathcal{O}})^2 \} + 6 (1 - \lambda_A^{\mathcal{O}})^2 \Tr \{ \Lambda_A^{\mathcal{O}} \gamma_A^{\mathcal{O}} (\Lambda_A^{\mathcal{O}})^T \gamma_A^{\mathcal{O}} \} \\
        &\hspace{0.4cm}+ 24 (1 - \lambda_A^{\mathcal{O}})\Tr \{ (\Lambda_A^{\mathcal{O}} \gamma_A^{\mathcal{O}})^2 (\Lambda_A^{\mathcal{O}})^T \gamma_A^{\mathcal{O}} \} + 8 (1 - \lambda_A^{\mathcal{O}}) \Tr \{ (\Lambda_A^{\mathcal{O}} \gamma_A^{\mathcal{O}})^3 \} \\
        &\hspace{0.4cm}+ 6 \Tr \{ (\Lambda_A^{\mathcal{O}} \gamma_A^{\mathcal{O}})^2 \} \Tr \{ \Lambda_A^{\mathcal{O}} \gamma_A^{\mathcal{O}} (\Lambda_A^{\mathcal{O}})^T \gamma_A^{\mathcal{O}} \} + 3 \Tr \{ \Lambda_A^{\mathcal{O}} \gamma_A^{\mathcal{O}} (\Lambda_A^{\mathcal{O}})^T \gamma_A^{\mathcal{O}} \}^2 \\
        &\hspace{0.4cm}+ 3 \Tr \{ (\Lambda_A^{\mathcal{O}} \gamma_A^{\mathcal{O}})^2 \}^2 + 24 \Tr \{ (\Lambda_A^{\mathcal{O}} \gamma_A^{\mathcal{O}})^3 (\Lambda_A^{\mathcal{O}})^T \gamma_A^{\mathcal{O}} \} + 6 \Tr \{ (\Lambda_A^{\mathcal{O}} \gamma_A^{\mathcal{O}})^4 \} \\
        &\hspace{0.4cm}+ 12 \Tr \{ (\Lambda_A^{\mathcal{O}} \gamma_A^{\mathcal{O}})^2 [(\Lambda_A^{\mathcal{O}})^T \gamma_A^{\mathcal{O}}]^2 \} + 6 \Tr \{ [\Lambda_A^{\mathcal{O}} \gamma_A^{\mathcal{O}} (\Lambda_A^{\mathcal{O}})^T \gamma_A^{\mathcal{O}}]^2 \},
    \end{split}
    \label{eq:GaussianIntegralsIIu4}
\end{equation}
which finally leads to
\begin{equation}
    \begin{split}
        U_A^{\mathcal{O}} (4) &= \left( \frac{1 + 3 \lambda_A^{\mathcal{O}}}{4} \right)^4 + \frac{3}{8} \left( \frac{1 + 3 \lambda_A^{\mathcal{O}}}{4} \right)^2 \Tr \left\{ \Lambda_A^{\mathcal{O}} \gamma_A^{\mathcal{O}} \left[ (\Lambda_A^{\mathcal{O}})^T + \Lambda_A^{\mathcal{O}} \right] \gamma_A^{\mathcal{O}} \right\} \\
        &\hspace{0.4cm}+ \frac{1}{8} \left(\frac{1 + 3 \lambda_A^{\mathcal{O}}}{4} \right) \Tr \left\{ (\Lambda_A^{\mathcal{O}} \gamma_A^{\mathcal{O}})^2 \left[ 3 (\Lambda_A^{\mathcal{O}})^T + \Lambda_A^{\mathcal{O}} \right] \gamma_A^{\mathcal{O}} \right\} + \frac{3}{256} \Tr \left\{ \Lambda_A^{\mathcal{O}} \gamma_A^{\mathcal{O}} \left[ (\Lambda_A^{\mathcal{O}})^T + \Lambda_A^{\mathcal{O}} \right] \gamma_A^{\mathcal{O}} \right\}^2 \\
        &\hspace{0.4cm}+ \frac{3}{128} \Tr \left\{ (\Lambda_A^{\mathcal{O}} \gamma_A^{\mathcal{O}})^2 \left[2 ((\Lambda_A^{\mathcal{O}})^T \gamma_A^{\mathcal{O}})^2 + (\Lambda_A^{\mathcal{O}} \gamma_A^{\mathcal{O}})^2 \right] \right\} \\
        &\hspace{0.4cm}+ \frac{3}{128} \Tr \left\{ \Lambda_A^{\mathcal{O}} \gamma_A^{\mathcal{O}} \left[ (\Lambda_A^{\mathcal{O}})^T + 4 \Lambda_A^{\mathcal{O}} \right] \gamma_A^{\mathcal{O}} \Lambda_A^{\mathcal{O}} \gamma_A^{\mathcal{O}} (\Lambda_A^{\mathcal{O}})^T \gamma_A^{\mathcal{O}} \right\}.
    \end{split}
    \label{eq:GaussianIntegralsIIU4}
\end{equation}

\subsubsection{Leading contributions for large particle energies}
As discussed in the main text, quasi-particle excitations are expected to exceed the ground-state entanglement when the particles carry sufficient energy. As we will show now, in this scenario the crucial building blocks of the Rényi entropies reduce to simple expressions that remain valid in the continuum. Exemplary, we consider the first term of the scalar $\lambda_{k,A}^f$, see Eq. \eqref{eq:QuasiParticleMarginalXi}, and take the continuum limit
\begin{equation}
    \begin{split}
        (\lambda_{k,A}^f)_1 &= \frac{1}{(\tilde{\gamma}_0^f)_{k k}} \sum_{j,j' \in \mathfrak{J}_B} \epsilon^2 \, e^{- i \epsilon (j-j') \eta k} (\gamma_{0,B}^f)_{j'j} \\
        &\to \omega_k \sum_{k'= - \infty}^{\infty} \left(\frac{\eta}{2 \pi} \right)^2 \frac{1}{\omega_{k'}} \int_l^L \mathrm{d}x \, e^{- i x \eta (k-k')} \int_l^L \mathrm{d}x' \, e^{-i x' \eta (k-k')} \\
        &= \frac{\omega_k}{\pi^2} \sum_{k'= - \infty}^{\infty} \frac{1}{\omega_{k'}} \frac{\sin^2 \left[ \eta \frac{L-l}{2} (k-k') \right]}{(k-k')^2}.
    \end{split}
\end{equation}
Note here that sums in position and momentum space can be interchanged by the Tonelli-Fubini theorem since the overall sum is absolutely convergent. To implement the approximation of large particle energies, we perform a variable transformation $k'' = \frac{L-l}{2} (k-k')$ (note that the sum over $k''$ runs in steps of $\frac{L-l}{2}$), such that 
\begin{equation}
    (\lambda_{k,A}^f)_1 = \frac{\omega_k}{\pi^2} \sum_{k''= - \infty}^{\infty}  \frac{L-l}{\sqrt{m^2 (L-l)^2 + \left[ \eta k (L-l) - 2 \eta k'' \right]^2}} \frac{(L-l)^2}{4} \frac{\sin^2 \left( \eta k'' \right) }{k''^2}.
    \label{eq:ParticleslambdaStep1}
\end{equation}
When either $m (L - l) \gg 1$ or $\eta k (L-l) \gg 1$, \textit{i.e.}, for massive or fast particles, the square root becomes independent of the running momentum variable $k''$, in which case the sum simplifies to
\begin{equation}
    (\lambda_{k,A}^f)_1 \to \frac{1}{\pi^2} \sum_{k''= - \infty}^{\infty} \frac{(L-l)^2}{4} \frac{\sin^2 \left( \eta k'' \right) }{k''^2}.
    \label{eq:ParticleslambdaStep2}
\end{equation}
The remaining sum can be evaluated after transforming back to the variable $k'$, resulting in the scaling 
\begin{equation}
    (\lambda_{k,A}^f)_1 = \pi^2 \sum_{k'= - \infty}^{\infty} \frac{\sin^2 \left[ \pi \left( 1 - \frac{l}{L} \right) (k-k') \right] }{(k-k')^2} = 1 - \frac{l}{L}.
\end{equation}
Performing an analogous calculation for the second term yields $ (\lambda_{k,A}^f)_2 \to 0$. Also, the momentum field expression is identical since the dispersion cancels out during the crucial step from Eq. \eqref{eq:ParticleslambdaStep1} to Eq. \eqref{eq:ParticleslambdaStep2}. This together with the phase-space relations Eqs. \eqref{eq:QuasiParticleWignerxi} and \eqref{eq:QuasiParticleHusimixiXi} (note that the prefactors appearing in the Husimi $Q$-expressions cancel out since $\lambda_{k,A}^{f}=\lambda_{k,A}^{g}$) gives simple scalings for particles carrying large energies, to wit
\begin{equation}
    \lambda_{k,A}^{f,g,\mathcal{Q}} \to  1 - \frac{l}{L}, \quad \lambda_{k,A}^{\mathcal{W}} \to 1 - 2 \frac{l}{L}.
    \label{eq:LambdasScaling}
\end{equation}
This analysis can be extended to the other terms appearing in $U_A^{\mathcal{O}} (r)$ involving $\Lambda_A^{\mathcal{O}}$. In general, traces over matrix products containing $\Lambda_A^{\mathcal{O}} \gamma_A^{\mathcal{O}}$ and $(\Lambda_A^{\mathcal{O}})^T \gamma_A^{\mathcal{O}}$ scale with the number of factors, \textit{i.e.},
\begin{equation}
    \Tr \{ \left(\Lambda_A^{\mathcal{O}} \gamma_A^{\mathcal{O}} \right)^n \} \propto \left( \frac{l}{L} \right)^n.
\end{equation}
The prefactors depend on the distribution $\mathcal{O}$ as well as on the number of matrices with (without) a transpose. After notoriously lengthy but straightforward calculations we find the overall scalings summarized in \autoref{tab:Us}.

\begin{table}[h!]
    \renewcommand{\arraystretch}{2}
    \setlength{\tabcolsep}{10pt}
    \begin{tabular}{ c | c | c | c } 
        $U_A^{\mathcal{O}} (r)$ & $r=2$ & $r=3$ & $r=4$ \\ 
        \hline
        $f,g$ & $1 - \frac{l}{L} + \frac{3}{4} \left( \frac{l}{L} \right)^2$ & $1 - 2 \frac{l}{L} + 2 \left( \frac{l}{L} \right)^2 - \frac{4}{9} \left( \frac{l}{L} \right)^3$ & $1 - 3 \frac{l}{L} + \frac{33}{8} \left( \frac{l}{L} \right)^2 - \frac{37}{16} \left( \frac{l}{L} \right)^3 + \frac{153}{256} \left( \frac{l}{L} \right)^4$ \\ 
        $\mathcal{W}$ & $1 - 2 \frac{l}{L} + 2 \left( \frac{l}{L} \right)^2$ & $1 - 4 \frac{l}{L} + \frac{20}{3} \left( \frac{l}{L} \right)^2 - \frac{32}{9} \left( \frac{l}{L} \right)^3$ & $1 - 6 \frac{l}{L} + 15 \left( \frac{l}{L} \right)^2 - 17 \left( \frac{l}{L} \right)^3 + \frac{15}{2} \left( \frac{l}{L} \right)^4$
    \end{tabular}
    \caption{$U_A^{\mathcal{O}} (r)$ for $r=2,3,4$ and the Wigner $W$- and marginal distributions in the limit of large particle energies.} 
    \label{tab:Us}
\end{table}
Calculating the Husimi $Q$-expressions is substantially more involved since the convolution introduces a dependence on the lattice spacing $\epsilon$. Hence, we only give the analytic result for $r=2$, which reads
\begin{equation}
    U_A^{\mathcal{Q}} (2) \to 1 - \frac{l}{L} + \frac{1}{4} \left[ 1 + \frac{2 - (5/6) \epsilon \omega_k + 2 (\epsilon \omega_k)^2}{(1 + \epsilon \omega_k)^2} \right] \left( \frac{l}{L} \right)^2.
    \label{eq:UHusimiQLimit}
\end{equation}
In the continuum limit $\epsilon \to 0$ and for large but finite momenta, the latter simplifies to
\begin{equation}
    U_A^{\mathcal{Q}} (2) \to 1 - \frac{l}{L} + \frac{3}{4} \left( \frac{l}{L} \right)^2,
\end{equation}
exemplifying that the expressions of the marginal distributions and the Husimi $Q$-distribution agree in the continuum, see the top-left cell of \autoref{tab:Us}. This extends to all values of $r$ since the prefactors appearing in Eq. \eqref{eq:QuasiParticleHusimixiXi2} vanish in the continuum limit except for the one associated with the field distribution $f$, see Eq. \eqref{eq:QuasiParticleHusimiQPrefactors}.

\subsubsection{Small intervals}
With the scalings \eqref{eq:LambdasScaling} at hand, we now additionally consider the case of small intervals $l \ll L$ to identify the first-order correction appearing for a quasi-particle with respect to the ground-state area law in the ultraviolet regime. To this end, we shall expand $U_A^{\mathcal{O}} (r)$ to first order in $l/L$, for which we need to expand $u_A^{\mathcal{O}} (s)$. For every entropic order $s \in \mathbb{N}$, there is precisely \textit{one} diagram where two consecutive vertices indices are connected to a closed loop (see first diagrams in \autoref{fig:Diagrams1}, \autoref{fig:Diagrams2}, \autoref{fig:Diagrams3} and \autoref{fig:Diagrams4}). Remarkably, this is the only diagram contributing to first order in $l/L$ since all other diagrams are at least of order $(l/L)^2$. More precisely, the aforementioned diagram gives rise to the term $(1 - \lambda_A^{\mathcal{O}})^s$, while the remaining diagrams involve at least \textit{two} matrices of the form $\Lambda_A^{\mathcal{O}} \gamma_A^{\mathcal{O}}$ under the trace. Hence, we have
\begin{equation}
    u_A^{\mathcal{O}} (s) \to (1 - \lambda_A^{\mathcal{O}})^s
\end{equation}
to leading order in $l/L$. Upon inspecting \eqref{eq:GaussianIntegralsIIGeneralPower}, this implies
\begin{equation}
    U_A^{\mathcal{O}} (r) \to \sum_{s=0}^r \begin{pmatrix}
            r \\ s
        \end{pmatrix} 
        \frac{(\lambda_A^{\mathcal{O}})^{r-s}}{r^s} \, (1 - \lambda_A^{\mathcal{O}})^s = \left[ \frac{1 + (r-1)\lambda_A^{\mathcal{O}}}{r} \right]^r.
\end{equation}
Given that $\lambda_A^{\mathcal{O}} \to 1 - a^{\mathcal{O}}_1 (l/L)$ with $a^{\mathcal{Q}}_1=a^{f}_1=a^{g}_1=1$ for the marginal and Husimi $Q$-distributions and $a^{\mathcal{W}}_1=2$ for the Wigner $W$-distribution, see Eq. \eqref{eq:LambdasScaling}, we end up with
\begin{equation}
    U_A^{\mathcal{O}} (r) \to 1 + a^{\mathcal{O}}_1 (1-r) \frac{l}{L} \equiv 1 + a^{\mathcal{O}}_{r,1} \frac{l}{L}
\end{equation}
to leading order in $l/L$, which is in agreement with \autoref{tab:Us} and Eq. \eqref{eq:UHusimiQLimit}. We remark that the latter holds for all entropic orders by analytic continuation.

\subsection{Subtracted Rényi entropies}

\subsubsection{Subtracting extensive contributions}
Given the decomposition \eqref{eq:GaussianIntegralsIIGeneralPower}, any Rényi entropy of the local distribution $\mathcal{O}_A$ corresponding to a Gaussian times a quadratic form can be written as
\begin{equation}
    S_r [\mathcal{O}_A] = \frac{1}{1-r} \ln \left[ \frac{\text{det}^{(1-r)/2} (2 \pi \gamma_A^{\mathcal{O}})}{r^{\dim (\gamma_A^{\mathcal{O}})/2}} \, U_A^{\mathcal{O}} (r) \right] \equiv \frac{1}{2} \ln \text{det}(2 \pi \gamma_A^{\mathcal{O}}) + \frac{1}{2} \frac{\ln r}{r-1} \dim (\gamma_A^{\mathcal{O}}) + \delta S_r [\mathcal{O}_A].
\end{equation}
Any vacuum distribution $\bar{\mathcal{O}}_A$ is of Gaussian form, such that $U_r^{\bar{\mathcal{O}}} = 1$, and thus its entropy is a function of the vacuum covariance matrix $\bar{\gamma}_A$ only, \textit{i.e.},
\begin{equation}
    S_r [\bar{\mathcal{O}}_A] = \frac{1}{2} \ln \text{det}(2 \pi \bar{\gamma}_A^{\mathcal{O}}) + \frac{1}{2} \frac{\ln r}{r-1} \dim (\bar{\gamma}_A^{\mathcal{O}}).
\end{equation}
Subtracting the latter from the former results in the general formula for the subtracted Rényi entropy presented in the main text,
\begin{equation}
    \Delta S_r [\mathcal{O}_A] = \frac{1}{2} \ln \text{det} \left[ \gamma_A^{\mathcal{O}} (\bar{\gamma}_A^{\mathcal{O}})^{-1} \right] +  \delta S_r [\mathcal{O}_A].
\end{equation}

\subsubsection{Wigner entropies}
First, we note that the inverse covariance matrix of the vacuum reads [cf. Eq. \eqref{eq:VacuumStateMarginalCovarianceNormalizationPositionSpace}]
\begin{equation}
    (\bar{\gamma}_A^{\mathcal{W}})^{-1}_{j j'} = \frac{1}{\epsilon} \begin{pmatrix}
        \frac{2}{\epsilon} & 0 \\
        0 & 2 \epsilon
    \end{pmatrix} \delta_{j j'}.
\end{equation}
We restrict to the special case for which the covariance matrix $\gamma_A^{\mathcal{W}}$ of the Gaussian part of $\mathcal{W}_A$ is a direct sum with respect to the marginal covariance matrices, which includes all three classes of states of our interest. Then, using that
\begin{equation}
    \sum_{j'} \epsilon \, (\gamma_A^{f})_{j j'} (\bar{\gamma}_A^{f})^{-1}_{j' j''} = \frac{2}{\epsilon} (\gamma_A^{f})_{j j''}, \quad  \sum_{j'} \epsilon \, (\gamma_A^{g})_{j j'} (\bar{\gamma}_A^{g})^{-1}_{j' j''} = 2 \epsilon (\gamma_A^{g})_{j j''},
    \label{eq:LocalMatrixProductInverse}
\end{equation}
results in
\begin{equation}
    \Delta S_r [\mathcal{W}_A] = \frac{1}{2} \ln \det \left( \frac{2}{\epsilon} \, \gamma_A^f \right) + \frac{1}{2} \ln \det \left( 2 \epsilon \, \gamma_A^g \right) + \delta S_r [\mathcal{W}_A].
    \label{eq:SubtractedRenyiWignerEntropy}
\end{equation}
Note again that the determinant in position space is defined with an extra $\epsilon$, see Eq. \eqref{eq:DiscreteDetTrPositionSpace}.

\subsubsection{Marginal entropies}
Based on \eqref{eq:LocalMatrixProductInverse}, we find for the marginal distributions
\begin{equation}
    \Delta S_r [f_A] = \frac{1}{2} \ln \det \left( \frac{2}{\epsilon} \, \gamma_A^f \right) + \delta S_r [f_A], \quad \Delta S_r [g_A] = \frac{1}{2} \ln \det \left( 2 \epsilon \, \gamma_A^g \right) + \delta S_r [g_A].
\end{equation}
Under the above assumption, the marginal expressions are related to the Wigner $W$-expression via
\begin{equation}
    \Delta S_r [\mathcal{W}_A] = \Delta S_r [f_A] + \Delta S_r [g_A] + \frac{1}{1-r} \ln \frac{U^{\mathcal{W}}_A (r)}{U_A^f (r) \, U_A^g (r)}.
\end{equation}

\subsubsection{Wehrl entropies}
Since $\bar{\gamma}_A^{\mathcal{O}} = 2 \bar{\gamma}_A^{\mathcal{W}}$ by definition, the matrix products of the marginal components evaluate to
\begin{equation}
    \sum_{j'} \epsilon \, (\gamma_A^{\mathcal{Q},f})_{j j'} (\bar{\gamma}_A^{\mathcal{Q},f})^{-1}_{j' j''} = \frac{1}{\epsilon} \, (\gamma_A^{\mathcal{Q},f})_{j j''}, \quad  \sum_{j'} \epsilon \, (\gamma_A^{\mathcal{Q},g})_{j j'} (\bar{\gamma}_A^{\mathcal{Q},g})^{-1}_{j' j''} = \epsilon \, (\gamma_A^{\mathcal{Q},g})_{j j''},
    \label{eq:LocalMatrixProductHusimiInverse}
\end{equation}
such that, together with \eqref{eq:HusimiQMarginalCovariance}, 
\begin{equation}
    \Delta S_r [\mathcal{Q}_A] = \frac{1}{2} \ln \det \left[ \frac{1}{\epsilon} \left( \gamma_A^f + \frac{1}{2} \right) \right] + \frac{1}{2} \ln \det \left[ \epsilon \left( \gamma_A^g + \frac{1}{2 \epsilon^2} \right) \right] + \delta S_r [\mathcal{Q}_A].
\end{equation}
Comparing with \eqref{eq:SubtractedRenyiWignerEntropy} reveals that the subtracted Wehrl entropy carries another extensive contribution coming from defining the Husimi $Q$-distribution via convolving the Wigner $W$-distribution. Therefore, we instead consider
\begin{equation}
    \Delta S_r [\mathcal{Q}_A] - (l/\epsilon) \ln 2 = \frac{1}{2} \ln \det \left[ \frac{2}{\epsilon} \left( \gamma_A^f + \frac{1}{2} \right) \right] + \frac{1}{2} \ln \det \left[ 2\epsilon \left( \gamma_A^g + \frac{1}{2 \epsilon^2} \right) \right] + \delta S_r [\mathcal{Q}_A].
\end{equation}

\section{Area laws for classical mutual informations}
\label{sec:AreaLawsMutualInformations}
The area law can be proven rigorously for classical mutual informations by formulating the argument in \cite{Wolf2008} for distributions instead of density operators. We consider the thermal state 
\begin{equation}
    \boldsymbol{\rho}_T = \frac{1}{Z_T} e^{-\beta \boldsymbol{H}}
\end{equation}
of some generic local Hamiltonian
\begin{equation}
    \boldsymbol{H} = \boldsymbol{H}_A + \boldsymbol{H}_B + \boldsymbol{H}_{\partial A},
    \label{eq:BipartiteHamiltoniansDecomposition}
\end{equation}
with the subsystem Hamiltonians $\boldsymbol{H}_A$ and $ \boldsymbol{H}_B$, and some local interaction $\boldsymbol{H}_{\partial A}$ (one may equally write $\boldsymbol{H}_{\partial B}$). Now, every distribution of our interest $\mathcal{O} [\nu]$ corresponds to a (positive) operator-valued measure $\boldsymbol{O}$. Hence, expressed via classical distributions, the thermal state reads
\begin{equation}
    \mathcal{O}_T [\nu] = \frac{1}{Z_T^{\mathcal{O}}} \, e^{- \beta H^{\mathcal{O}}},
\end{equation}
with the corresponding classical Hamiltonian $H^{\mathcal{O}}$ being defined implicitly via
\begin{equation}
    e^{- H^{\mathcal{O}}} = \Tr \{e^{- \boldsymbol{H}} \, \boldsymbol{O} \}.
\end{equation}
In this case, the classical Rényi entropies can be written as
\begin{equation}
    S_r [\mathcal{O}_T] = \frac{r}{r-1} \ln Z_T^{\mathcal{O}} + \frac{1}{1-r} \ln \left[ \int \mathcal{D}\nu \, e^{- r \beta H^{\mathcal{O}}} \right] \equiv \frac{r}{r-1} \ln Z_T^{\mathcal{O}} + \braket{H^{\mathcal{O}}}_r^{\text{ln}},
    \label{eq:ClassicalEntropyThermal}
\end{equation}
where
\begin{equation}
    \braket{H^{\mathcal{O}}}_r^{\text{ln}} = \frac{1}{1-r} \ln \left[ \int \mathcal{D}\nu \, e^{- r \beta H^{\mathcal{O}}} \right]
\end{equation}
denotes the logarithmic average of order $r$. The latter can be interpreted as a generalization of the energy expectation value, which is obtained in the special case $r \to 1$, \textit{i.e.}, 
\begin{equation}
    \braket{H^{\mathcal{O}}}^{\text{ln}}_1 = \int \mathcal{D}\nu \, \mathcal{O}_T [\nu] \, \beta H^{\mathcal{O}} = \beta \braket{H^{\mathcal{O}}}.
\end{equation}
Next, we introduce the classical free energy and express it in terms of classical Rényi entropies, to wit
\begin{equation}
    F [\mathcal{O}_T] = - \frac{\ln Z_T^{\mathcal{O}}}{\beta} = \frac{1}{\beta} \frac{1-r}{r} \left\{ \braket{H^{\mathcal{O}}}_r^{\text{ln}} - S_r [\mathcal{O}_T] \right\}.
    \label{eq:ClassicalFreeEnergyThermal}
\end{equation}
We note that although most terms on the right-hand side of the latter equation depend on the entropic order $r$, the left-hand side is independent of $r$, and hence the equation holds for all $r$. Now, it is well-known that free energy serves as the thermodynamic potential for the canonical ensemble and as such attains its minimum precisely for the canonical state, \textit{i.e.}, $F [\mathcal{O}_T] \le  F [\mathcal{O}]$. This holds true in particular when we disregard correlations between $A$ and $B$ by considering the product distribution $\mathcal{O}_{T,A} [\nu_A] \times \mathcal{O}_{T,B} [\nu_B]$, which corresponds to the product state $\boldsymbol{\rho}_{T,A} \otimes \boldsymbol{\rho}_{T,B}$, \textit{i.e.},
\begin{equation}
    F [\mathcal{O}_T] \le F [\mathcal{O}_{T,A} \times \mathcal{O}_{T,B}].
\end{equation}
Using \eqref{eq:ClassicalFreeEnergyThermal} together with the additivity of classical entropies leads to
\begin{equation}
    I_r [\mathcal{O}_{T,A} : \mathcal{O}_{T,B}] = S_r [\mathcal{O}_{T,A}] + S_r [\mathcal{O}_{T,B}] - S_r [\mathcal{O}_T] \le \braket{H^{\mathcal{O}_A \times \mathcal{O}_B}}_r^{\text{ln}} - \braket{H^{\mathcal{O}}}_r^{\text{ln}}.
\end{equation}
However, the decomposition \eqref{eq:BipartiteHamiltoniansDecomposition} implies that the generalized energy expectation values of $\mathcal{O}_T$ and $\mathcal{O}_{A,T} \times \mathcal{O}_{B,T}$ agree in the two subregions $A$ and $B$. Therefore, they cancel out on the right-hand side of the latter inequality, which leaves us with
\begin{equation}
    I_r [\mathcal{O}_{T,A} : \mathcal{O}_{T,B}] \le \braket{H_{\partial A}^{\mathcal{O}_A \times \mathcal{O}_B}}_r^{\text{ln}} - \braket{H_{\partial A}^{\mathcal{O}}}_r^{\text{ln}}.
\end{equation}
Since the remaining bound is a function of the classical boundary Hamiltonian $H_{\partial}$ only, every classical Rényi mutual information does at most scale with the surface area $\abs{\partial A}$ of the boundary region up to some proportionality constant.

\section{Supplementary figures}
\label{sec:Figures}

\subsection{Central charge and ground-state area laws}
\label{subsec:CentralCharge}
\begin{figure*}[h!]
    \centering
    \includegraphics[width=0.99\textwidth]{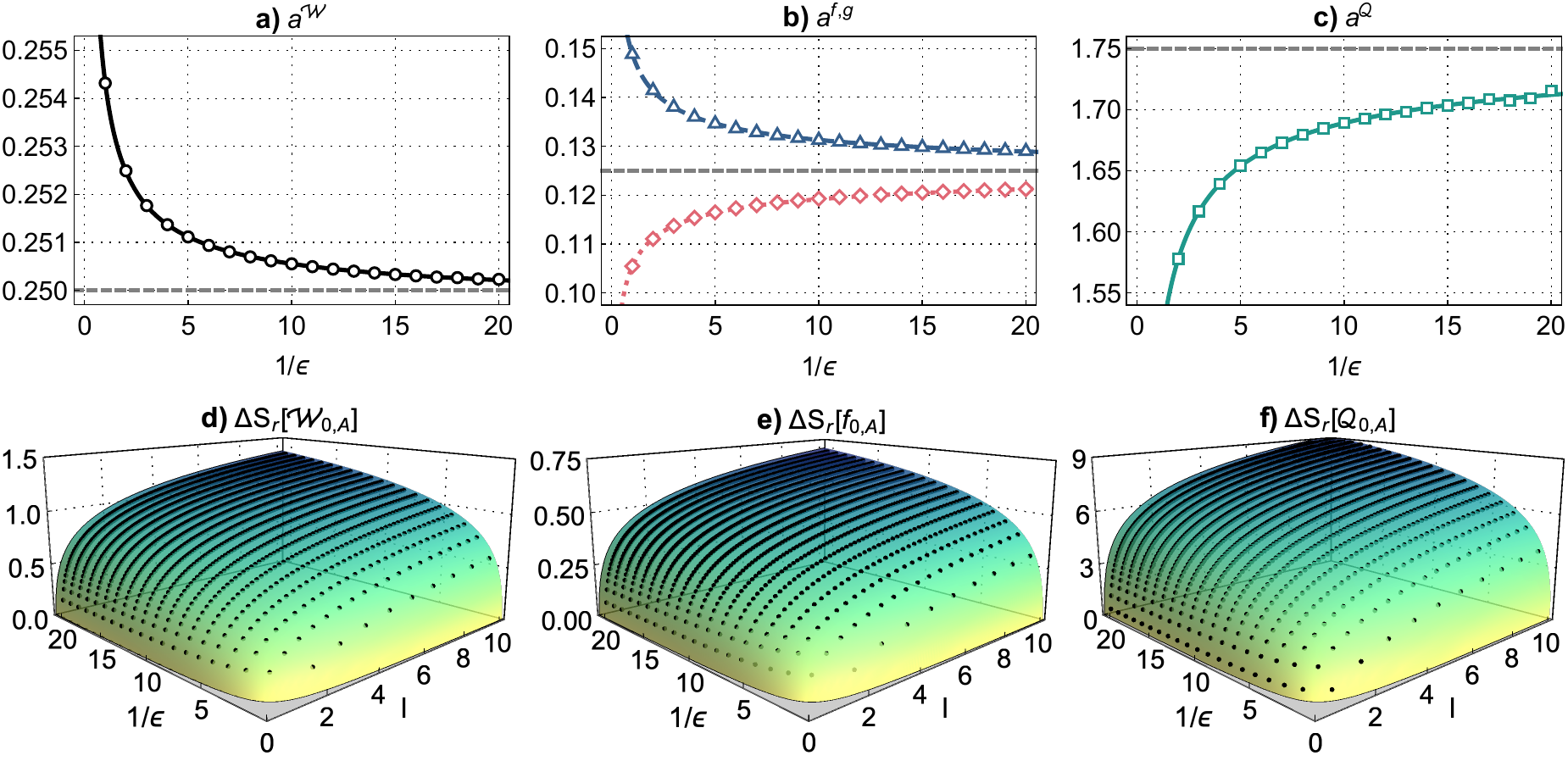}
    \caption{Extended numerical analysis of the ground-state area law in the conformal theory ($m = 10^{-10}, L=200 \gg l=10$). In the upper row, we plot the prefactor $a^{\mathcal{O}}$ appearing in the logarithmic area-law fit $a^{\mathcal{O}} \ln (l / \epsilon) + b^{\mathcal{O}}$ to the numerically evaluated subtracted entropies $\Delta S [\mathcal{O}_{0,A}]$ for decreasing lattice spacing $\epsilon = \{ 1, 1/2, ..., 1/20 \}$ (points). The curves correspond to rational fits of the form $c_1 + c_2 \epsilon^{c_3}$, such that $c_1$ serves as an approximation to the continuum value of $a^{\mathcal{O}}$ (gray dashed lines). All four curves indicate convergence toward the continuum limit. That the area-law fits (colored contours) accurately fit the numerical data (black points) is confirmed in the lower row. While the subtracted Wigner and marginal entropies are known to fulfill area laws due to their relations to the Rényi-2 entanglement entropy, we also find an area law for the subtracted Wehrl entropy $\Delta S_r [\mathcal{Q}_{0,A}] - (l/\epsilon) \ln 2$, see \textbf{f)}.}
    \label{fig:EntropiesGSSM}
\end{figure*}

It is well-known that the prefactor $a$ of the logarithmic area law of the ground state is proportional to the central charge $c$ of the theory. Here, we provide further numerical evidence on this connection for the three types of classical entropies of our interest from the lattice perspective. In the lattice theory, the prefactor $a$ is extracted by fitting the area law to the numerically obtained data points. This procedure introduces a regularization dependence in the sense that $a = a(\epsilon, m, L)$, which we wish to analyze toward the continuum limit $\epsilon \to 0$. To this end, we consider the conformal theory on the real line and set $m = 10^{-10}, L=200 \gg l=10$. We show the dependence of the prefactor $a$ for all four distributions as a function of the ultraviolet regulator $1/\epsilon$ in \autoref{fig:EntropiesGSSM} \textbf{a)} - \textbf{c)}. We test the overall validity of the area law fit, in particular, the dependence of the data on $\sim \ln (l / \epsilon)$ only, by fitting the latter to \textit{all} data points, see \autoref{fig:EntropiesGSSM} \textbf{d)} - \textbf{f)}.

\subsection{Thermal correlations}
\label{subsec:CorrelationsThermalState}
\begin{figure*}[h!]
    \centering
    \includegraphics[width=0.65\textwidth]{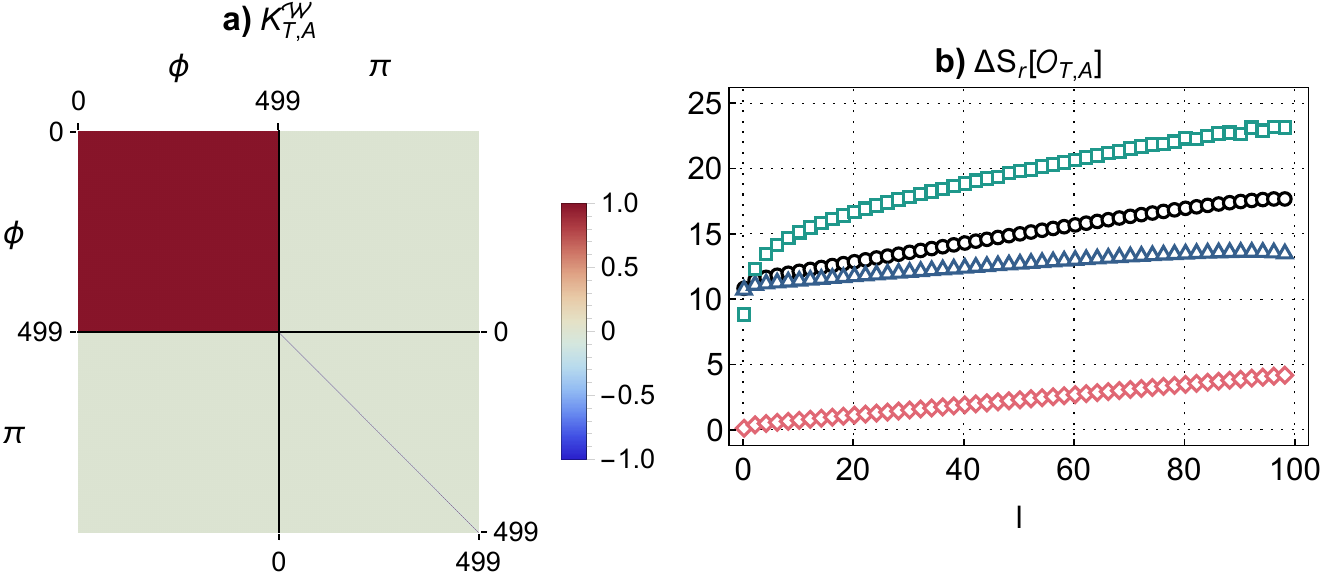}
    \caption{\textbf{a)} Wigner correlation matrix of the thermal state ($\epsilon = 0.2, m = 10^{-6}, L=100, T=10^{-1}$). The correlations are mostly encoded in the field. \textbf{b)} Subtracted entropies for the same scenario (without offsets subtracted). The entropy associated with $\pi$ (red) is less sensitive to finite system size than the entropy of $\phi$ (blue points) when $l \to L$, thereby leading to a smaller mutual information for the momentum field.}
    \label{fig:EntropiesTSSM}
\end{figure*}

To develop a deeper understanding of why thermal correlations are mostly encoded in the field $\phi$, see \textcolor{blue}{Figure 2} \textbf{c)} in the main text, we consider the correlation matrix with coefficients
\begin{equation}
    (K^{\mathcal{O}}_{A})_{j j'} = \frac{(\gamma_A^{\mathcal{O}})_{j j'}}{\sqrt{(\gamma_A^{\mathcal{O}})_{j j} \, (\gamma_A^{\mathcal{O}})_{j' j'}}} \in [-1,1],
\end{equation}
which allows for a visual assessment of the strength of correlations. We show the correlation matrix of the Wigner $W$-distribution for the scenario of \textcolor{blue}{Figure 2} \textbf{c)} in the main text in \autoref{fig:EntropiesTSSM} \textbf{a)}. The upper left block describing field correlations contains elements very close to $1$, while the non-diagonal elements of the lower right block describing momentum field correlations are negligible (the off-diagonal blocks are strictly zero as the thermal Wigner $W$-distribution is a product distribution with respect to the two marginal $f$ and $g$).

Further, we show all subtracted entropies up to the total system size and without offsets subtracted in \autoref{fig:EntropiesTSSM} \textbf{b)}. It appears that the entropy over the momentum field is generally small compared to the entropy of the field itself. More importantly, the curves differ substantially when $l \to L$, see $l \approx 95$: while the entropy of $\phi$ (blue points) decreases noticeably when finite system size dominates over finite temperature, the entropy of $\pi$ (red points) is substantially less sensitive to finite-size effects. As a result, the corresponding mutual information curve is overall flatter since the subtracted entropy of the full system ($l=L=10^2$) is subtracted no matter how the system is partitioned.

\subsection{Subtracted Rényi entropies of a quasi-particle}
\label{subsec:HigherOrderRenyiEntropies}
\begin{figure*}[h!]
    \centering
    \includegraphics[width=0.99\textwidth]{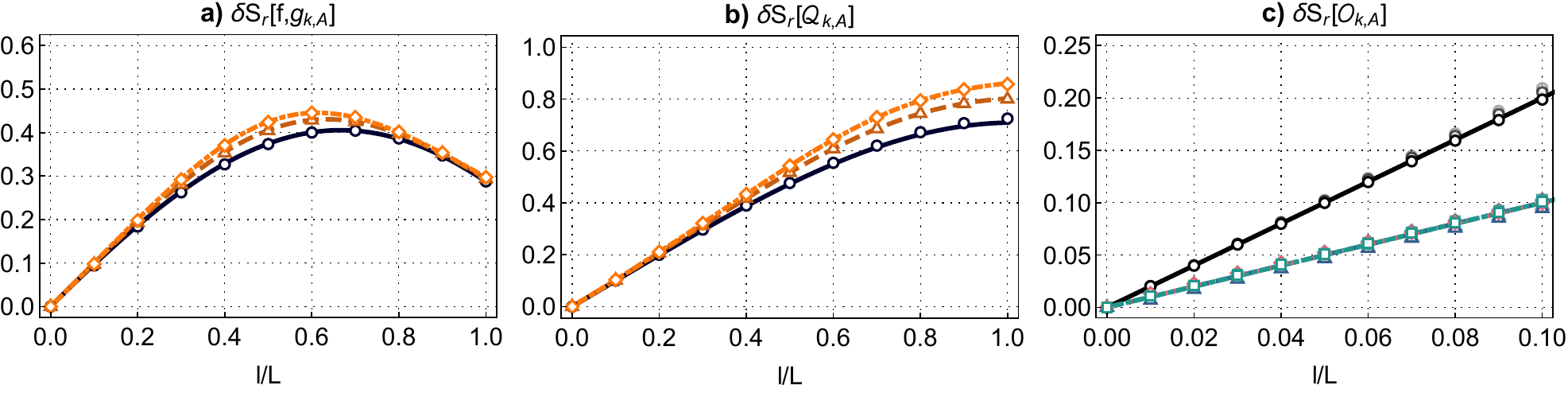}
    \caption{Entropy surplus of a fast and massive quasi-particle ($k=50, m=10, L = 10, \epsilon=10^{-1}$). \textbf{a)} and \textbf{b)} complement \textcolor{blue}{Figure 2} \textbf{e)} and show the subtracted Rényi entropies of order $r=2,3,4$ (black straight, dashed brown, dotted orange) for the marginal and Husimi $Q$-distribution, respectively. While the curves in \textbf{a)} are obtained using the scalings summarized in \autoref{tab:Us}, we use the analytic result \eqref{eq:UHusimiQLimit} for $r=2$ in \textbf{b)} and interpolations otherwise. Locally, all subtracted Rényi entropies scale linearly with $(2) l/L$, which we exemplify in \textbf{c)} for $r=2,3,4$ up to $l/L=0.1$.}
    \label{fig:EntropiesParticlesSM}
\end{figure*}

In analogy to \textcolor{blue}{Figure 2} \textbf{e)} of the main text, we show the subtracted marginal and Wehrl entropies of the quasi-particle state for the three entropic orders $r=2,3,4$ in \autoref{fig:EntropiesParticlesSM} \textbf{a)} and \textbf{b)}, respectively. Additionally, we plot \textit{all} considered subtracted Rényi entropies in the regime of small intervals $l \ll L$ in \autoref{fig:EntropiesParticlesSM} \textbf{c)}, showing that the entropy surplus of a quasi-particle is linear to leading order and independent of the entropic order.

\bibliography{references.bib}